\documentclass[fleqn,usenatbib]{mnras}
\usepackage{comment}
\usepackage{setspace}
\usepackage{braket}
\usepackage{url}
\usepackage{longtable}
\usepackage{pdflscape}

\usepackage{newtxtext,newtxmath}

\usepackage[T1]{fontenc}
\usepackage{ae,aecompl}

\usepackage{graphicx}	
\usepackage{amsmath}	
\usepackage{amssymb}	

\title[ diffuse molecular gases in NGC~1300]{A large amount of diffuse molecular gases in the bar of the strongly barred galaxy NGC~1300: Cause of the low star formation efficiency}

\author[F. Maeda et al.]{
Fumiya Maeda,$^{1}$\thanks{E-mail: fmaeda@kusastro.kyoto-u.ac.jp}
Kouji Ohta,$^{1}$
Yusuke Fujimoto,$^{2}$
Asao Habe,$^{3}$
and Kaito Ushio$^{1}$
\\
$^{1}$Department of Astronomy, Kyoto University, Kitashirakawa-Oiwake-Cho, Sakyo-ku, Kyoto, Kyoto 606-8502, Japan\\
$^{2}$Earth and Planets Laboratory, Carnegie Institution for Science, 5241 Broad Branch Road, NW, Washington, DC 20015, USA\\
$^{3}$Graduate School of Science, Hokkaido University, Kita 10 Nishi 8, Kita-ku, Sapporo, Hokkaido 060-0810, Japan
}

\date{Accepted XXX. Received YYY; in original form ZZZ}

\pubyear{2019}

\begin{document}
\label{firstpage}
\pagerange{\pageref{firstpage}--\pageref{lastpage}}
\maketitle

\begin{abstract}
In many barred galaxies, star formation efficiency (SFE) in the bar is lower than those in the arm and bar-end, and its cause has still not been clear.
Focusing on the strongly barred galaxy NGC~1300, we investigate the possibility that the presence of a large amount of diffuse molecular gas, which would not contribute to the SF, makes the SFE low in appearance. We examine the relation between the SFE and the diffuse molecular gas fraction ($f_{\rm dif}$), which is derived using  the $^{12}$CO($1-0$) flux obtained from the interferometer of ALMA 12-m array, which has no sensitivity on diffuse (extended; FWHM $\gtrapprox 700$ pc) molecular gases due to the lack of ACA, and the total $^{12}$CO($1-0$) flux obtained from Nobeyama 45-m single-dish telescope. We find that the SFE decreases with increasing $f_{\rm dif}$. The $f_{\rm dif}$ and ${\rm SFE}$ are $0.74 - 0.91$ and  $(0.06 - 0.16) ~\rm Gyr^{-1}$ in the bar regions, and  $0.28 - 0.65$ and $(0.23 - 0.96) ~\rm Gyr^{-1}$ in the arm and bar-end regions. This result supports the idea that the presence of a large amount of diffuse molecular gas makes the SFE low. The suppression of the SFE in the bar has also been seen even when we exclude the diffuse molecular gas components.
This suggests that the low SFE appears to be caused not only by a large amount of diffuse molecular gases but also by other mechanisms such as fast cloud--cloud collisions. 
\end{abstract}




\section{Introduction} \label{sec: Intro}
It has been commonly known that massive star formation (SF) in bars of barred galaxies is suppressed. In the radial profile of H$\alpha$ line emission,  which is one of the massive SF tracers,
there is a strong dip near the bar regions of early-type barred galaxies \citep{James2009A&A}. In galaxies with bars of intermediate strength, the SF efficiency (SFE $= \Sigma_{\rm SFR}/\Sigma_{\rm H_2}$) in the bar region is 2 $\sim$ 3 times lower than that in the arm region (e.g. M83; \citealt{Hirota:2014bt}, NGC~4303; \citealt{Momose2010ApJ}). In strongly barred galaxies, the absence of massive SF along the bars is clearly seen \citep[e.g. NGC~1300 and NGC 5383;][]{Tubbs1982ApJ}.
Prominent H\textsc{ii} regions are often not seen in the bar regions, even though there are remarkable dust lanes along the stellar bar where the amounts of the observed molecular gas are similar to those in other galactic regions
\citep[e.g.][]{Downes1996ApJ, Maeda:2018bg}.
Another evidence for the massive SF suppression was provided from an independent analysis by \citet{Hakobyan2016MNRAS}.
They found that the radial distribution of core-collapse supernovae in barred galaxies is different from that of unbarred galaxies, while the distributions of Type Ia supernovae are not significantly different. These results can be explained by the suppression of massive SF in the radial range of the bars.
However, physical mechanisms which suppress the massive SF in the bar regions are still unclear.

This question is a long-standing problem, and several explanations have been proposed in observational and theoretical studies. 
\citet{Tubbs1982ApJ} made numerical hydrodynamic simulations of a strongly barred galaxy, NGC 5383 to test the destruction of dense gas clouds in the gravitational field by the strong bar. \citet{Tubbs1982ApJ} suggested the molecular clouds may be destroyed by shock due to the high velocity of the gases relative to the bar structure.
Hydrodynamic simulations by \citet{Athanassoula1992MNRAS} showed that the density of the gas in the bar region is low, except for the loci of shocks. Along the shocks, 
there is a high shear which probably suppresses the molecular cloud formation \citep[see also][]{Zurita2004A&A}. 
The role of shear for the suppression of SF within the bar regions is also suggested in recent Milky Way-like simulation at sub-pc resolution \citep{Emsellem2015MNRAS}.
CO observations towards NGC~1530 suggest intense shock with high-velocity jumps and a large shear suppress the SF by destroying the molecular clouds \citep{ReynaudDownes1998A&A}.

Recent studies proposed other scenarios and physical mechanisms that could suppress the SF fall into the following three scenarios.
The first scenario is that molecular clouds in the bar regions may be gravitationally unbound. 
\citet{Sorai2012PASJ} made $^{12}$CO($1-0$) map of Maffei 2 at an angular resolution of 200 pc with a single-dish telescope,
and pointed out a possibility that clouds in the bar are gravitationally unbound, which causes the low SF activity \citep[see also][]{Meidt2013ApJ}.
\citet{Nimori2013MNRAS} performed a 2D hydrodynamical simulation and also found the unbound clouds in the bar regions.

The second scenario is that 
fast cloud--cloud collisions (CCCs) may occur in the bar regions. CCCs induce clump formation by shock compressions and have been suggested as the mechanism of massive SF \citep[e.g.][]{Habe1992PASJ,Fukui2014ApJ...780,Renaud2015MNRAS}. \citet{Fujimoto:2014eb} performed a high resolution ($\sim$ a few pc) 3D hydrodynamical simulation of an intermediate-type barred galaxy and found that collision velocity between the clouds in the bar regions is larger than that in the arm regions. Based on this simulation, 
\citet{Fujimoto:2014kh} proposed 
that the fast CCC of the clouds in the bars shortens a gas accretion
phase of the cloud cores formed, leading to suppression of core growth and massive SF \citep[see also][]{Takahira:2014cq,Takahira:2018gx}.
Note that the intermediate and small stars can form in the fast CCCs.
Recent observations towards the central region of the Milky Way detected a fast CCC without massive SF   \citep{Enokiya2019PASJ}.

The third scenario is that a large amount of diffuse molecular gas may exist in the bar regions. Molecular gas traced by $^{12}$CO($1-0$) would consist of two components: One is giant molecular clouds (GMCs).
The mass and size of the GMC are  $\sim 10^{5-6}~M_\odot$ and $\sim 10 - 100~\rm pc$ \citep[e.g.][]{Solomon87}, and the SF occurs in GMCs.  
The other is the extended diffuse component which is distributed on scales larger than sub-kpc.
Presence of such diffuse component  is reported by
\citet{Pety:2013fw} and \citet{Caldu-Primo2015AJ}; they found that about half of the $^{12}$CO($1-0$) emission arises mostly from spatial scales larger than 1.3~kpc in M51 and 1.0~kpc in NGC~4736, respectively.
In the SFE calculation using $^{12}$CO($1-0$) emission, the diffuse molecular gas, which would not directly contribute to the current SF activity, is included in the $\Sigma_{\rm H_2}$.
The third scenario proposes that the diffuse gas greatly contributes to the $\Sigma_{\rm H_2}$ in the bar regions and that the SFE becomes apparently low in appearance. 
Single-dish CO observations towards barred galaxies show gas density in the bar regions is lower than that in the arm regions using non-LTE analysis (e.g. NGC~2903; \citealt{Muraoka:2016ip}, and NGC~4303; \citealt{Yajima:2019do}).
This suggests that a large amount of diffuse molecular gases exists in the bar. In the Milky Way, a dense gas fraction in the Galactic bar region is smaller than that in the Galactic arm region, which would imply the presence of a large amount of diffuse gases in the bar \citep{Torii2019PASJ}. However, it is still unclear which scenario is the most dominant cause for the low SFE in the bar regions.

\begin{figure}
	\includegraphics[width=85mm]{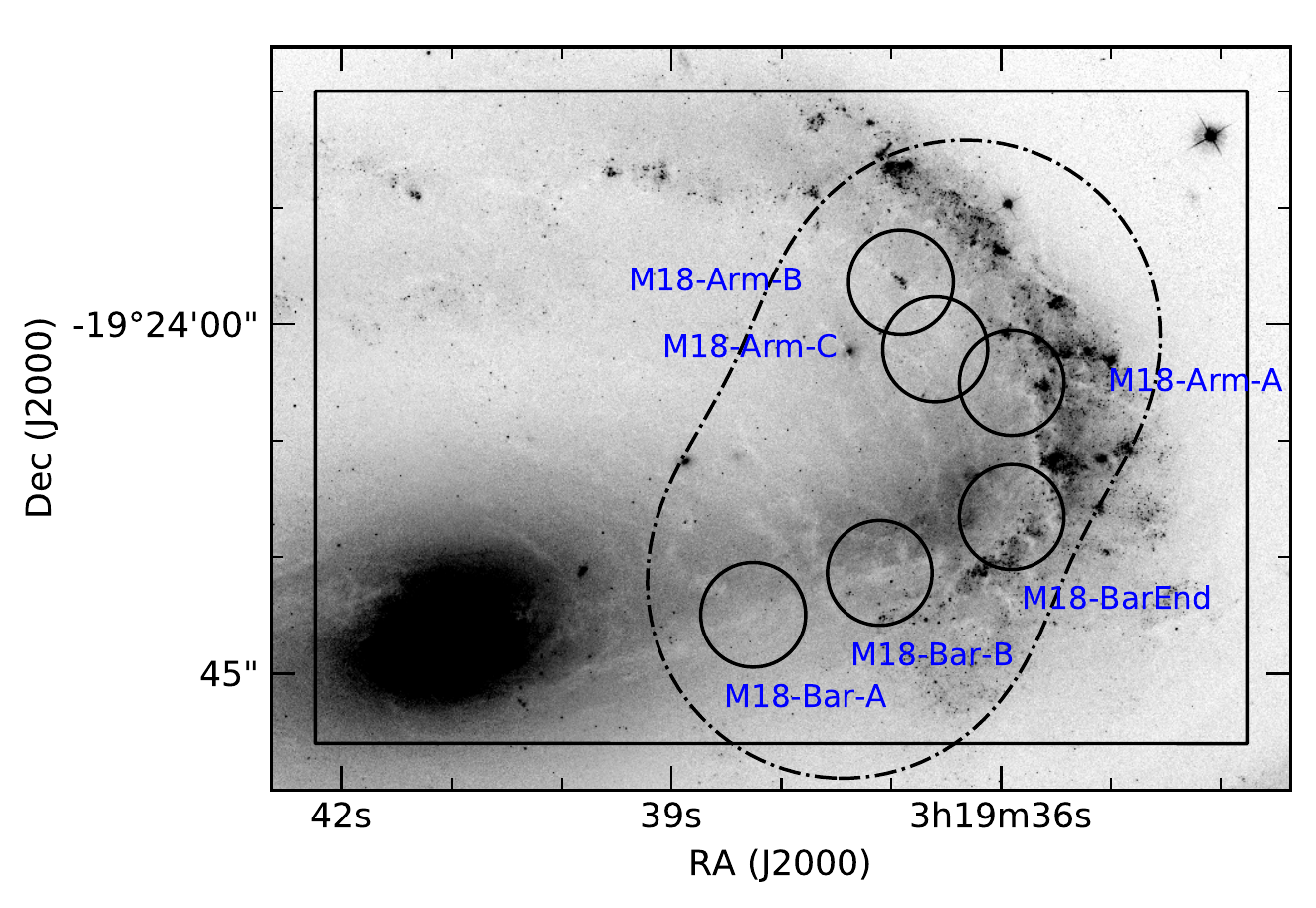}
    \caption{$V$-band image of NGC~1300 taken with {\it F555W} filter on Advanced Camera for Surveys (ACS) of the Hubble Space Telescope (HST).   We obtained this image from the Hubble Legacy Archive (HLA; \url{https://hla.stsci.edu/}). The black rectangle represents the mapping area observed with the NRO 45-m (NRO FoV). The black dash--dotted line represents the FoV observed with ALMA \citep{Maeda2020MNRAS} (ALMA FoV). 
    The black solid circles represent the observed positions with the previous single-pointing observations with the NRO 45-m telescope \citep{Maeda:2018bg}.
    The diameter of the circles show the beam size of 13.5 arcsec of the NRO 45-m.
    }
 \label{fig:NRO_FoV}
\end{figure}

In order to understand the cause for the suppression of the SF in the bar regions, the strongly barred galaxy NGC~1300 is one of the suitable laboratories because the absence of SF is clearly seen in the bar regions; remarkable dust lanes are seen in the bar regions without prominent H\textsc{ii} regions, while in the arm regions H\textsc{ii} regions are associated with the dust lanes (Fig.~\ref{fig:NRO_FoV}). Thus,
physical mechanism(s) of the suppression is(are) expected to be clearly seen. 
$^{12}$CO($1-0$) observations at a high angular resolution of $\sim40$ pc with Atacama Large Millimeter/submillimeter Array (ALMA) towards NGC~1300 showed that there is no clear difference in the distribution of the virial parameter, which is a measure for gravitational binding of molecular clouds \citep{BertoldiandMckee}, between the GMCs in the bar region and those in the arm region \citep{Maeda2020MNRAS}. This result suggests that the lack of massive SF in the strong bar of NGC~1300 can not be explained by a systematic difference of the virial parameter (the first scenario).
\citet{Fujimoto2020MNRAS} presented a hydrodynamical simulation of a strongly barred galaxy, using a stellar potential model of NGC~1300. They found that there is no significant environmental dependence of cloud properties including the virial parameter, which is qualitatively consistent with the results of the observations by \citet{Maeda2020MNRAS}. Further, they showed that the collision speed in the bar is significantly faster than the other regions due to the elongated global gas motion by the stellar bar. They concluded that fast CCCs (the second scenario) would be one of the  physical mechanisms for the SF suppression.

\begin{table}
 \caption{Adopted parameters of NGC~1300}
 \label{tab:NGC1300}
 \begin{tabular}{lc}
  \hline
  Parameter & Value \\
  \hline
  Morphology$^a$ & SB(s)bc \\
  Centre position (J2000.0)$^b$ & $\rm 03^h19^m41^s.036$  \\
  & $\rm -19^\circ24^\prime40^{\prime\prime}.00$\\
  Inclination$^c$ & $50^\circ.2$ \\
  Distance$^d$ & 20.7 Mpc \\
  Linear scale & 100 $\rm pc~arcsec^{-1}$ \\
\hline
\multicolumn{2}{l}{{\small$^a$  \citet{Sandae_Tammann}}} \\
\multicolumn{2}{l}{{\small $^b$  The peak of V-band image as Fig. ~\ref{fig:NRO_FoV}}}\\
\multicolumn{2}{l}{{\small$^c$  \citet{England1989a}}}\\
\multicolumn{2}{l}{{\small$^d$ We adopted the systemic velocity with corrections for }}\\
\multicolumn{2}{l}{{\small  the Virgo cluster, the Great Attractor, and the Shapley}}\\
\multicolumn{2}{l}{{\small  concentration of $1511~{\rm km~s^{-1}}$ \citep{MouldEtAl00} and }}\\
\multicolumn{2}{l}{{\small  the Hubble constant of $73~{\rm km~s^{-1}~Mpc^{-1}}$. }}
 \end{tabular}
\end{table}

In this paper, we investigate the third scenario that a large amount of diffuse molecular gases may exist in the bar regions. 
Diffuse molecular gases in the Milky Way are well studied through emission lines of CO isotopes and absorption lines \citep[e.g.][]{Snow2006ARA&A,Sheffer2008ApJ,Liszt&Pety2012A&A,Roman-Duval2016ApJ}. 
Because observations of such lines towards nearby galaxies are difficult, we use
the recovery fraction  of the CO flux ($f_{\rm re}$), 
which is the ratio of CO flux obtained from interferometers to total CO flux obtained from single-dishes, as an alternative measure of the amount of the diffuse molecular gases. 
Interferometers provide higher spatial resolution with a disadvantage that not all spatial scales can be recovered.
The shortest baselines determine the largest spatial scale which the interferometers can recover. Therefore, the recovery fraction 
shows the percentage of the molecular gases that is more compact than the largest scale. Recovery fraction was measured in some galaxies and the presence of diffuse molecular gases is suggested (e.g. M51; \citealt{Pety:2013fw}, NGC~4736 and NGC~5055; \citealt{Caldu-Primo2015AJ}). However, the difference in the recovery fraction with environments and the relation between the recovery fraction and SFE have not been adequately studied.

Instead of $f_{\rm re}$, we present the diffuse gas fraction, $f_{\rm dif}$, defined as $1 - f_{\rm re}$ in NGC~1300 using ALMA $^{12}$CO($1-0$) observations of \citet{Maeda2020MNRAS}, which has no sensitivity on extended molecular gases due to lack of Atacama Compact Array (ACA; 7-m + TP) data, and the total CO flux observations with the 45-m single-dish telescope of Nobeyama Radio Observatory (NRO).
We derive the SFR from archival H$\alpha$, FUV and IR data, 
which are the massive SF tracers, then investigate the relation between $f_{\rm dif}$ and SFE. Further, using the archival data of $^{12}$CO($2-1$) observations with ACA, we investigate the relation among the $f_{\rm dif}$, SFE, and the emission line ratio of the $^{12}$CO($2-1$) to $^{12}$CO($1-0$) ($R_{21/10}$), which can be a measure to know the physical condition of molecular gases \citep[e.g.][]{Koda2012ApJ761}.

This paper is structured as follows: In Section~\ref{sec:Data}, we describe our $^{12}$CO($1-0$) observations with the NRO 45-m (Section~\ref{sec: NRO CO(1-0)}) and with ALMA (Section~\ref{sec: ALMA CO(1-0)}).
We also summarize archival $^{12}$CO($2-1$), H$\alpha$, FUV and IR data used in our analysis (Section \ref{sec: Archival data}). Then, Section~\ref{sec: Conversion to physical parameters} 
presents the resultant molecular gas surface density and SFR. Results of stacking analysis also are presented in Section~\ref{sec: Result Stacking Analysis}. The $f_{\rm dif}$ and SFE in NGC~1300, which are  main results of the paper, are presented in Section~\ref{sec: Star formation efficiency and recovery fraction}. In Section~\ref{sec:discussion}, we discuss the cause for the SF suppression. Our conclusions are presented in Section~\ref{sec:summary}.
Appendix~\ref{apx} shows the supplementary results of our analysis. 
Table~\ref{tab:NGC1300} summarizes parameters of NGC~1300 adopted throughout this paper, which is the same as those in \citet{Maeda2020MNRAS}.

\section{Observations and data reduction} \label{sec:Data}

In this section, we describe making procedure of fits images we use in this paper; $^{12}$CO($1-0$) emission line obtained from the NRO 45-m (Section~\ref{sec: NRO CO(1-0)}), $^{12}$CO($1-0$) and $^{12}$CO($2-1$) emission lines from ALMA (Section~\ref{sec: ALMA CO(1-0)} and \ref{sec: ALMA CO(2-1)}), H$\alpha$ emission from HST (Section~\ref{sec: HST Ha}), FUV from GALEX, and 22$\mu$m from WISE (Section~\ref{sec: GALEX WISE}). 
The spatial resolution and pixel size of all images we use for analysis are convolved and regrided to those of the $^{12}$CO($1-0$) image obtained from the NRO 45-m, which have the poorest beam size of  $16.7$ arcsec and the pixel size of $6.0~{\rm arcsec} \times 6.0~{\rm arcsec}$.
In this paper, "pixel" refers to the region of $6.0~{\rm arcsec} \times 6.0~{\rm arcsec}$, unless otherwise noted.

\begin{figure}
	\includegraphics[width=85mm]{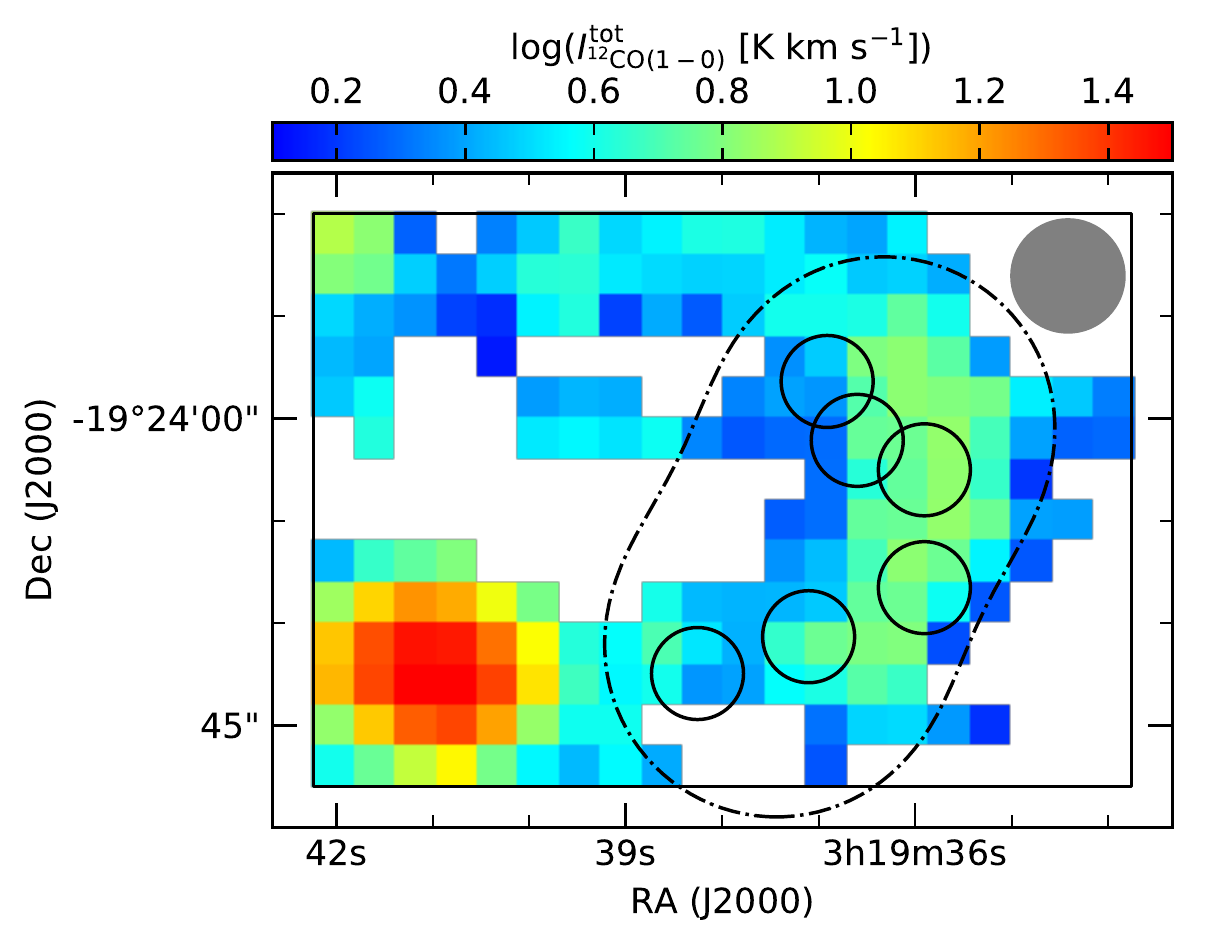}
    \caption{Velocity-integrated intensities map of $^{12}$CO($1-0$) line with the NRO 45-m (i.e. moment zero map). In white pixels, we did not detect any significant emissions. The grey circle represents the effective angular resolution of 16.7 arcsec. Other lines are the same as Fig.~\ref{fig:NRO_FoV} }
 \label{fig:Result_ICO10_NRO}

	\includegraphics[width=75mm]{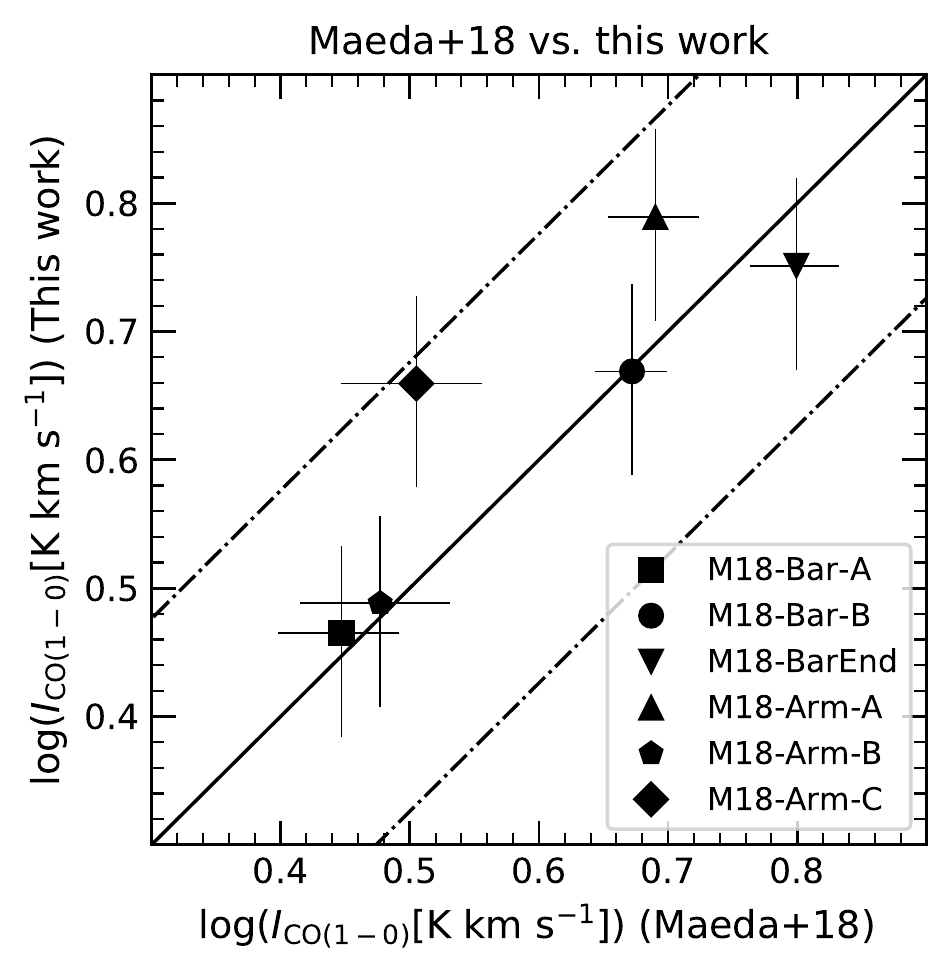}
    \caption{Comparison between $^{12}$CO($1-0$) line intensity obtained from this work and those by \citet{Maeda:2018bg}. 
    The legends represent the name of the observed regions, which are shown as black circles in Fig.~\ref{fig:NRO_FoV}, defined in  \citet{Maeda:2018bg}.
    The solid line and dash--dotted lines represent the 1:1 correlation and factor of 1.5 differences, respectively.}
 \label{fig:Maeda18_vs_Maeda20}
\end{figure}

\begin{figure*}
	\includegraphics[width=160mm]{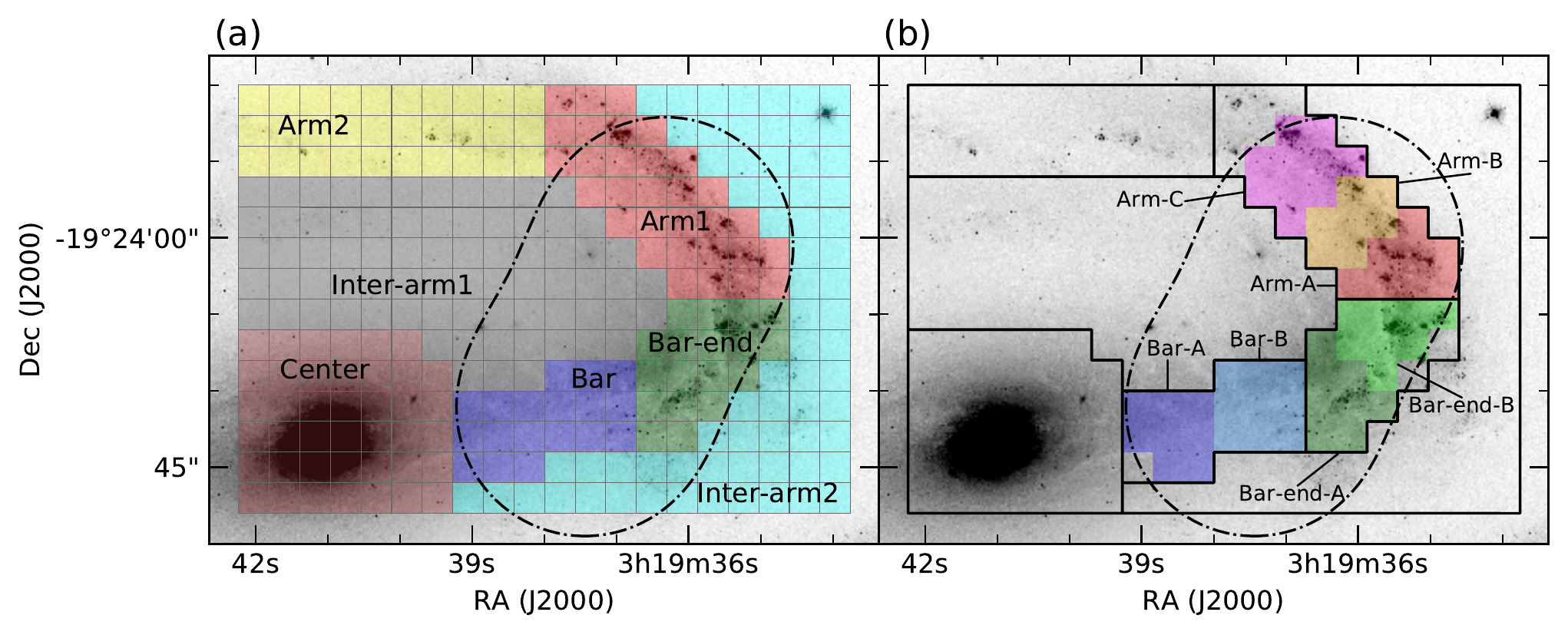}
    \caption{(a) Definition of the environmental mask for NRO FoV indicated with colour. The {\it centre}, {\it bar}, {\it bar-end}, {\it arm1}, {\it arm2}, {\it inter-arm1}, and {\it inter-arm2} are indicated with brown, blue, green, red, yellow, grey and cyan, respectively.
    The grey lines shows the $6.0~{\rm arcsec} \times 6.0~{\rm arcsec}$ grid.
    (b) Definition of the environmental mask for ALMA FoV indicated with colour.
    The {\it bar-A}, {\it bar-B}, {\it bar-end-A}, {\it bar-end-B}, {\it arm-A}, {\it arm-B}, and {\it arm-C} are indicated with blue, lightblue, green, lightgreen, red, orange and magenta, respectively.
    The black solid lines represent definition of the environmental mask for NRO FoV as panel (a). The black dash--dotted line represent the FoV observed with ALMA \citep{Maeda2020MNRAS}.}
 \label{fig:definition_of_environments}
\end{figure*}

\subsection{CO($1-0$) emission from the NRO 45-m observations} \label{sec: NRO CO(1-0)}

\subsubsection{Observations and data reduction}\label{sec: NRO CO(1-0), obs and reduction}
We carried out simultaneous observations of $^{12}$CO($1-0$) (rest frequency: 115.271202~GHz) and 
$^{13}$CO($1-0$) (110.201353~GHz) emission lines towards NGC~1300 on 2019 February 16, 17, 18, and 20
with the NRO 45-m,
employing the on-the-fly (OTF) mapping mode.
The observed area is $120~{\rm arcsec} \times 84~{\rm arcsec}$, corresponding to $12.0~{\rm kpc} \times 8.4~{\rm kpc}$, as shown in a black rectangle in Fig.~\ref{fig:NRO_FoV} (hereafter, we call this region NRO FoV).
This region covers from the centre to the western arm of NGC~1300, which also covers the regions observed with ALMA
(\citealt{Maeda2020MNRAS}; Section~\ref{sec: ALMA CO(1-0)}).

We used the multi-beam receiver, FOur-beam REceiver System on the 45-m Telescope \citep[FOREST;][]{Minamidani2016}.
FOREST has four beams, which are arranged on the corners of a $50$ arcsec square. Each beam is a dual-polarization sideband-separating SIS mixer receiver.
The beam size of each beam is $\sim 14$ arcsec at 115~GHz. The Intermediate Frequency (IF) bandwidth of 8~GHz enables us to simultaneously observe $^{12}$CO($1-0$) and  $^{13}$CO($1-0$).
The backend is an FX-type correlator system, SAM45, which consists of 16 arrays with 4096 spectral channels each. Two correlators are assigned for each beam and polarization.
The bandwidth and resolution were set to 1~GHz and 488.28~kHz, respectively, corresponding to $2600~\rm km~s^{-1}$ and $1.3~\rm km~s^{-1}$ at 115~GHz.
Although C$^{18}$O($1-0$) (109.782173~GHz) emission line is also within the bandwidth,
we do not focus on the line because its emission is expected to be extremely weak \citep[e.g.][]{Sorai2019PASJ}.

We performed the OTF mapping along the long side (X scan) and the short side (Y scan) of the black rectangle in Fig.~\ref{fig:NRO_FoV}.
The separation between the scan rows was set to $5.0$ arcsec,
and scan speeds were $8.8~\rm arcsec~s^{-1}$ for X scan and  $7.2~\rm arcsec~s^{-1}$ for Y scan, respectively. 
The pointing accuracy was checked almost every hour to keep within 3 arcsec by observing SiO maser source, $o$ Cet ( $22^\circ$ from NGC~1300), using a 43 GHz band receiver. 
The line intensity was calibrated by the chopper wheel method.
The system temperature ($T_{\rm sys}$) was 300–500~K at 115~GHz, which is slightly high due to the low elevation of $20^\circ - 30^\circ$.
The total on--source time was about 15 hrs.

The observed data were analyzed using the software package NOSTAR \citep{Sawada2008PASJ}, which comprises tools for OTF data analysis developed by NRO.
First, we divided each scan into five spectra and flagged spectra with poor baselines or/and spurious lines by eye.
Then, a baseline (second-order polynomial function with small curvature) 
was subtracted, and the data were regridded to 6.0 arcsec per pixel with an effective angular resolution of approximately 16.7 arcsec,
corresponding to 1.67 kpc. We smoothed the spectrum by binning to $20~\rm km~s^{-1}$.
Because the image rejection ratios (side band ratio) in the adopted frequency range were almost lager than 10 dB, no correction was made. 
We converted the antenna temperature ($T_{\rm A}^\ast$) into the main beam brightness temperature ($T_{\rm mb}$) using the main beam efficiency of 
$35 \pm 3$ percent, which is observatory-provided value for 2018-19 observing season. The resultant rms noise is 13.9 mK and 7.7 mK for $^{12}$CO($1-0$) and $^{13}$CO($1-0$), respectively, at $20~\rm km~s^{-1}$ bin.

\subsubsection{CO emissions}\label{sec: NRO CO(1-0), CO emission}
Using the data cube obtained,
we identified the significant CO emission.
First, we identified the channel bins in which signals are above $3\sigma_{\rm rms}$ in the spectrum of each pixel. Here, $\sigma_{\rm rms}$ is the rms noise of the spectrum.
Then, we expanded these channel bins to include all adjacent bins in which signals are above $\sigma_{\rm rms}$.
The velocity-integrated intensity, $I_{\rm ^{12}CO(1-0)}^{\rm tot}$, of each pixel is defined as sum of the identified channel bins.
Fig.~\ref{fig:Result_ICO10_NRO} shows the $^{12}$CO($1-0$)  map (i.e. moment zero map).
We show the pixels where significant $^{12}$CO($1-0$) emission was detected.
We detected the $^{12}$CO($1-0$) emission from the centre to arm including the bar and bar-end regions. 
$I_{\rm ^{12}CO(1-0)}^{\rm tot}$ is $10 \sim 15~\rm K~km~s^{-1}$ in centre and $3.0 \sim 6.0~\rm K~km~s^{-1}$ between the bar and arm.
However, we did not detect any significant $^{13}$CO($1-0$) emission lines.

We estimate the uncertainty of $I_{\rm ^{12}CO(1-0)}^{\rm tot}$ as $\sigma_{\rm ^{12}CO(1-0)}^{\rm tot} = \sqrt{N} \sigma_{\rm rms} \Delta V $, where $N$ is the number of channels (bins) used in the integration and $\Delta V$ is the channel width of $20~\rm km~s^{-1}$. 
The typical uncertainty is 17~percent.
An upper limit of the  $I_{\rm ^{12}CO(1-0)}^{\rm tot}$ where we did not detect any significant emission is estimated as $3\sigma_{\rm ^{12}CO(1-0)}^{\rm tot}$ by assuming $N = 2.5$, which corresponds to the mean FWHM of the $^{12}$CO($1-0$) emission lines observed by \citet{Maeda:2018bg}.

\subsubsection{Comparison with \citet{Maeda:2018bg}} \label{sec: NRO CO(1-0), comparison}
To see the accuracy of the  $I_{\rm ^{12}CO(1-0)}^{\rm tot}$ measurements,
we compared the $^{12}$CO($1-0$) line intensities obtained from this work with those by \citet{Maeda:2018bg}, which presented the single--pointing observations of several regions with the NRO 45-m.
The black circles in Fig.~\ref{fig:NRO_FoV} show the observed positions of  \citet{Maeda:2018bg} with the beam size (HPBW) of 13.5 arcsec of the NRO 45-m.
Using the moment zero map of Fig.~\ref{fig:Result_ICO10_NRO}, we measured the CO($1-0$) line intensity in the circle assuming the axially symmetric Gaussian beam pattern and  17~percent uncertainty of the line intensity, which is derived in Section~\ref{sec: NRO CO(1-0), CO emission}.
Fig.~\ref{fig:Maeda18_vs_Maeda20} shows the one--to--one relation of the $^{12}$CO($1-0$) line intensity.
The solid line and dash--dotted lines represent the 1:1 correlation and factor of 1.5 differences, respectively.
For the M18-Bar-A, M18-Bar-B, M18-Bar-End and M18-Arm-B regions, we find that the line intensity obtained from this work is mostly the
same as those by \citet{Maeda:2018bg}.
The line intensity in the M18-Arm-A and M18-Arm-C is different from \citet{Maeda:2018bg} by a factor of 1.3 and 1.5, respectively.
This result indicates that the $I_{\rm ^{12}CO(1-0)}^{\rm tot}$ measurements described in Section~\ref{sec: NRO CO(1-0), obs and reduction} are generally accurate but may contain systematic errors up to a factor of 1.5.

\subsubsection{Definition of environments} \label{sec: NRO CO(1-0), definition of environments}
In this study, we measure the physical parameters ($\Sigma_{\rm mol}$, $f_{\rm dif}$, $R_{21/10}$, SFR, and SFE) for each $6.0~{\rm arcsec} \times 6.0~{\rm arcsec}$ region, which is the pixel size of the $^{12}$CO($1-0$) data cube obtained from the NRO 45-m observations. The 6.0 arcsec grid in NRO FoV is shown in Fig.~\ref{fig:definition_of_environments}(a) and we separated these pixels into seven environments: {\it centre}, {\it bar}, {\it bar-end}, {\it arm1}, {\it arm2}, {\it inter-arm1}, and {\it inter-arm2}, according to colour regions in Fig.~\ref{fig:definition_of_environments}(a).
Considering the spread of $^{12}$CO($1-0$) emission in Fig.~\ref{fig:Result_ICO10_NRO}, we defined {\it centre} as a brown region.
The blue region is defined as {\it bar}, which covers the dark lane and associated spurs that are connected almost perpendicularly to the dark lane.
The green region that covers the intersection region of the bar and the arm is defined as {\it bar-end}.
The arm region is separated into {\it arm1} and {\it arm2} shown as red and yellow regions, respectively. While the bright H\textsc{ii} regions are seen in 
{\it arm1}, there are several H\textsc{ii} regions in {\it arm2}.
The remaining region is separated into {\it inter-arm1} and {\it inter-arm2} shown as grey and cyan region, respectively.
The environmental mask definition for the NRO FoV are only used to make the H$\alpha$ image in Section~\ref{sec: HST Ha} and to investigate the Kennicutt--Schmidt law and the relation between the SFE and $R_{21/10}$ in Appendix \ref{apx}.

Because the main purpose of this paper is to figure out the relation between the $f_{\rm dif}$ and SFE, we mainly use the pixels within the FoV of ALMA $^{12}$CO($1-0$) observations shown as the black dash--dotted line in Fig.~\ref{fig:NRO_FoV} (hereafter, we call ALMA FoV).
In order to compare the relation between the $f_{\rm dif}$ and SFE 
among galactic environments in details, we separated the pixels in {\it bar}, {\it bar-end} and {\it arm1} within ALMA FoV into  
{\it bar-A}(blue) and {\it bar-B}(light-blue), {\it bar-end-A}(green) and {\it bar-end-B}(light-green), and {\it arm-A}(red), {\it arm-B}(orange), and {\it arm-C}(magenta), respectively, according to Fig.~\ref{fig:definition_of_environments}(b).

\begin{figure}
	\includegraphics[width=70mm]{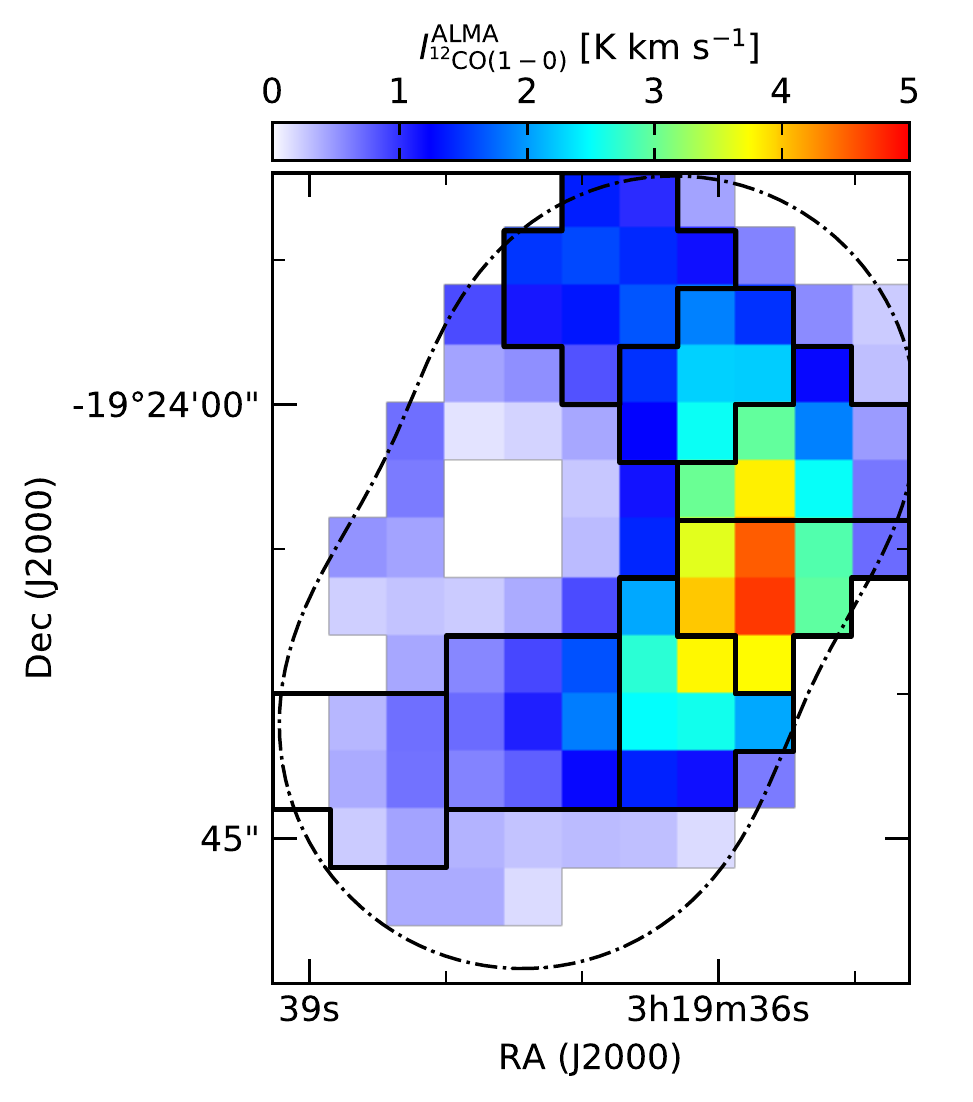}
    \caption{Velocity-integrated intensities map of $^{12}$CO($1-0$) line with ALMA convolved to the common beam size of 16.7 arcsec.
    The black solid lines represent definition of the environmental mask for ALMA FoV as Fig.~\ref{fig:definition_of_environments}(b).
    The black dash--dotted line represent the FoV observed with ALMA \citep{Maeda2020MNRAS}. }
 \label{fig:Result_ICO10_ALMA}

\begin{center}
	\includegraphics[width=70mm]{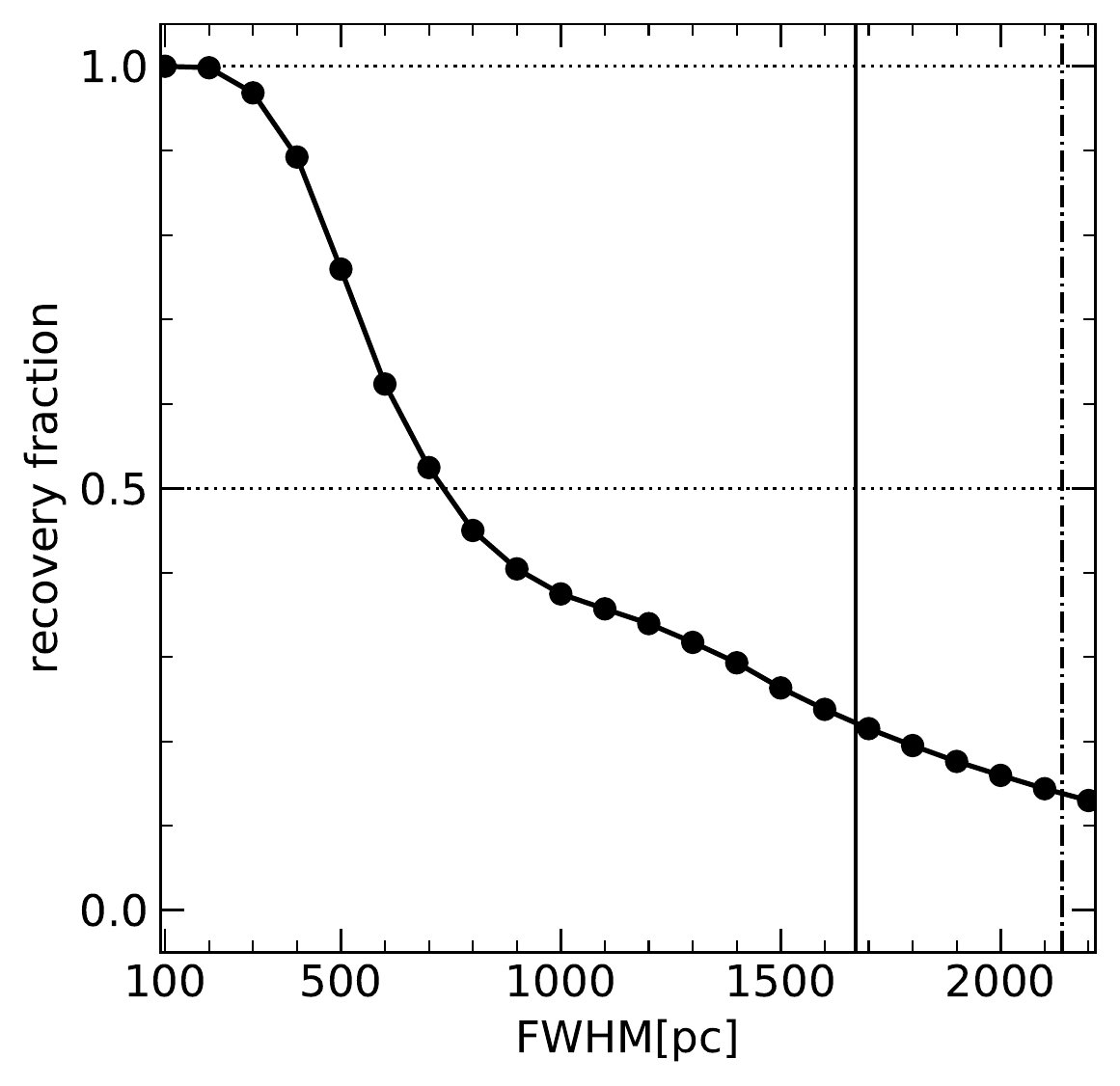}
    \caption{Recovery fraction of the Gaussian component as a function of the FWHM. We simulated ALMA observation of the mock Gaussian component with a given FWHM under the same configuration of the $^{12}$CO($1-0$) ALMA observations and noise-free condition. The horizontal dashed lines indicate the recovery fraction of 1.0 and 0.5.
    The vertical solid line indicates 1670 pc and the dash–dotted line indicates 2140 pc, corresponding to the beam size of the NRO 45-m telescope and, MRS of the $^{12}$CO($1-0$) ALMA observations, respectively.} 
 \label{fig:recovery_fraction_simulation}
 \end{center}
\end{figure}

\subsection{CO($1-0$) emission from ALMA observations} \label{sec: ALMA CO(1-0)}

\subsubsection{Observations and data reduction} \label{sec: ALMA CO(1-0), obs and reduction}
The details of the observations of $^{12}{\rm CO}(1-0)$ line with ALMA and data reduction of them are described in Section 2 in \citet{Maeda2020MNRAS}. Thus we give a brief summary here.
We carried out $^{12}{\rm CO}(1-0)$ line observations of NGC~1300 in Cycle 5 under project 2017.1.00248.S (PI = F. Maeda).
The observed region is shown as a black dash--dotted line in Fig.~\ref{fig:NRO_FoV}, where two pointings were set. The total on-source time was 5.31 hours (2.65 hours for each position).
We used about 44 antennas with C43-5 configuration in which the projected baseline length ranged from  15.1 m to 2.5 km, which corresponds to a maximum recoverable scale (MRS) of $\sim 21.4$ arcsec at 115 GHz. We used the Band 3 receiver with the central frequency of 114.664 GHz, channel width of 244.1 kHz ($\sim 0.64~\rm km~s^{-1}$), and bandwidth of 468.8 MHz ($\sim 1225~\rm km~s^{-1}$). 

Raw visibility data was calibrated by using the Common Astronomy Software Applications (\textsc{casa}) ver. 5.1.1. and the observatory-provided calibration script. We reconstructed the two-field mosaic image using \textsc{casa} ver. 5.4.0 using the \verb|multiscale| CLEAN algorithm \citep{Cornwell2008}
with Briggs weighting with robust = 0.5. 
We chose a velocity resolution of $5~\rm km~s^{-1}$. The resultant rms noise is 0.51 $\rm mJy~beam^{-1}$, corresponding to 0.36~K. We applied the primary beam correction on the output restored image and we extracted the region within the primary beam correction factor smaller than 2.0. The data cube has an angular resolution of $0.44~{\rm arcsec} \times 0.30~{\rm arcsec}$ and a pixel size of 0.12 arcsec.

Then, we convolved this data cube to the common beam size of 16.7 arcsec.
To identify the significant CO emission, we first identify the 3D (position-position-velocity) regions with a signal-to-noise ratio (S/N) $\geq$ 4 in at least two adjacent velocity channels.
We expand this mask to include all adjacent pixels with S/N $\geq 2$ in 3D space.
Finally, we regrided this data cube to 6.0 arcsec grid.
Fig.~\ref{fig:Result_ICO10_ALMA} shows the map of the velocity-integrated $^{12}$CO($1-0$) intensities ($I_{\rm ^{12}CO(1-0)}^{\rm ALMA}$). As the uncertainty of $I_{\rm ^{12}CO(1-0)}^{\rm ALMA}$ in each pixel, we consider the absolute flux calibration accuracy of $\pm 5$ percent in Band 3 (ALMA Technical Handbook).

\subsubsection{Missing flux} \label{sec: ALMA CO(1-0), missing flux}
Our ALMA observations did not recover the total $^{12}$CO($1-0$) flux due to the lack of ACA (7-m+TP) measurements.
Here, we check the relationship between the spatial scale of the gas distribution and missing flux, we simulated ALMA observation of a mock Gaussian component.
We created a fits image of a circular Gaussian component with a total flux of 1.0 Jy and given FWHM. The FWHM was set from 100 pc to 2200 pc in 100 pc interval. 
In order to extract the effect originated from the uv-distribution,
we simulated the observation under the same configuration and noise-free condition by using the task of \verb|simobserve| in \textsc{casa}. 
After reconstructing the image, we measured the flux of the component and recovery fraction. Fig.~\ref{fig:recovery_fraction_simulation} shows the recovery fraction of the Gaussian component as a function of the FWHM.
We find that the recovery fraction of the Gaussian component with FWHM $\leq 300$ pc is $\sim 1.0$ and that with FWHM $> 300$ pc is under 1.0. For the large Gaussian component with the FWHM $\geq 700$ pc, more than half of the flux is missed.
Thus, a molecular gas structure homogeneously extended to over $\sim 700$ pc is mostly resolved out in this $^{12}$CO($1-0$) ALMA observations. If FWHM is larger than the MRS of ALMA observation (2.14 kpc), flux recovers less than 10 percent.

\subsection{Archival data} \label{sec: Archival data}
\subsubsection{CO($2-1$) emission from ALMA observations } \label{sec: ALMA CO(2-1)}
\begin{figure}
	\includegraphics[width=85mm]{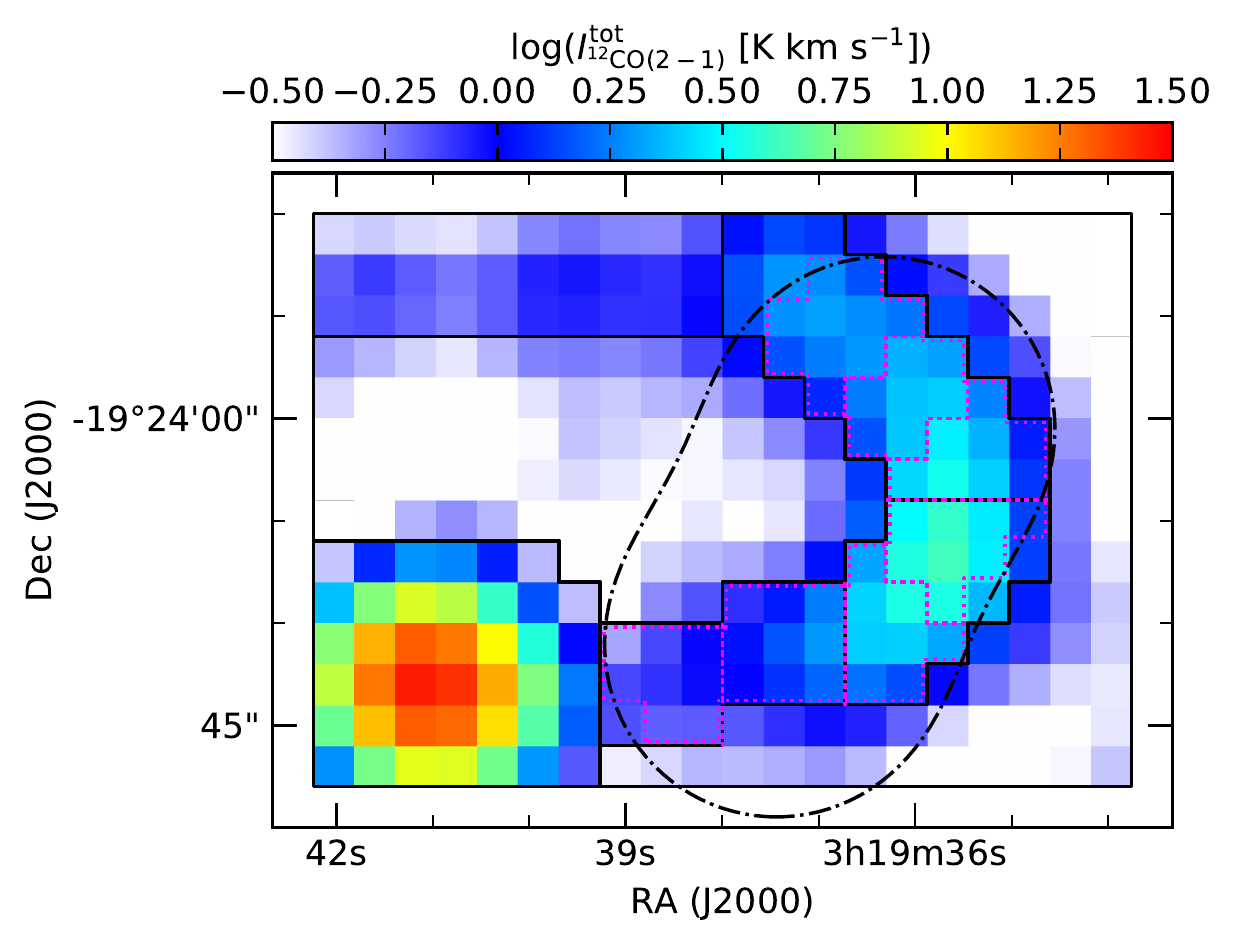}
    \caption{
    Velocity-integrated intensities map of $^{12}$CO($2-1$) line with ALMA convolved to the common beam size of 16.7 arcsec.
    The black solid lines and magenta dotted lines represent definition of the environmental mask for NRO FoV  and ALMA FoV as Fig.~\ref{fig:definition_of_environments}(a) and (b), respectively.
    The black dash--dotted line represent the FoV observed with ALMA \citep{Maeda2020MNRAS}.}
 \label{fig:Result_ICO21}
\end{figure}

We made a map of $^{12}$CO($2-1$) (rest frequency: 230.538000~GHz) emission using the archival data which was observed with ACA (7-m + TP) under project 2015.1.00925.S as proposed by B. Guillermo et al.
Thanks to the TP data, the total $^{12}$CO($2-1$) flux was detected.
We calibrated raw visibility data using \textsc{casa} and the observatory-provided calibration script. We imaged the interferometric map using the CLEAN algorithm in \textsc{casa} by adopting Briggs weighting with robust $= 0.5$. The resulting image is feathered with the total power image to recover the extended emission. 
The spatial resolution is $7.6~{\rm arcsec} \times 4.1~{\rm arcsec}$, corresponding to $760~{\rm pc} \times 410~{\rm pc}$. 
The rms noise of the data cube is $11.0~\rm mJy~beam^{-1}$ per 10.0 $\rm km~s^{-1}$ bin.
Then, we convolved this data cube to the common beam size of 16.7 arcsec and identified the significant $^{12}$CO($2-1$) emission as the same procedure described in Section~\ref{sec: ALMA CO(1-0), obs and reduction}. 
Finally, we regrided this data cube to 6.0 arcsec grid.
Fig.~\ref{fig:Result_ICO21} shows the map of the velocity-integrated $^{12}$CO($2-1$) intensities ($I_{\rm ^{12}CO(2-1)}^{\rm tot}$). As the uncertainty of $I_{\rm ^{12}CO(1-0)}^{\rm ALMA}$ in each pixel, we consider the absolute flux calibration accuracy of $\pm 10$ percent in Band 6 (ALMA Technical Handbook).

\begin{figure*}
	\includegraphics[width=\hsize]{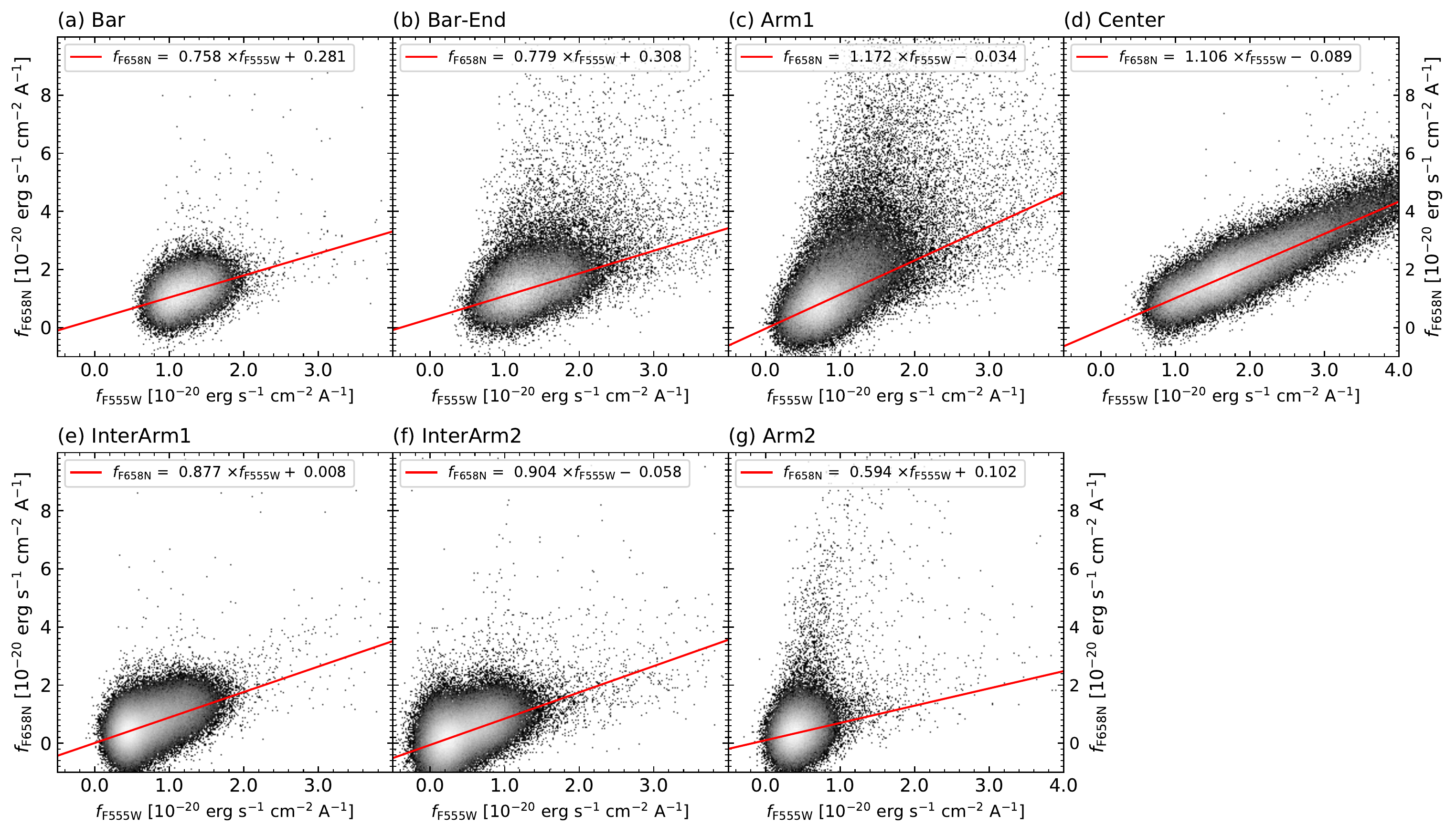}
    \caption{(a) The flux densities of the pixels in the {\it F658N} against those of the same pixel in the {\it F555W} image in {\it bar-end}. The red solid line shows the best-fit straight line to the data points by using robust least-squares regression (see text). 
    The best fit values are shown in top.
    (b)-(g) Same as panel (a) but for {\it centre}, {\it bar}, {\it arm1}, {\it arm2}, {\it inter-arm1}, and {\it inter-arm2}, respectively.} 
 \label{fig:continuum_fitting_result}
\end{figure*}

\begin{figure*}
	\includegraphics[width=160mm]{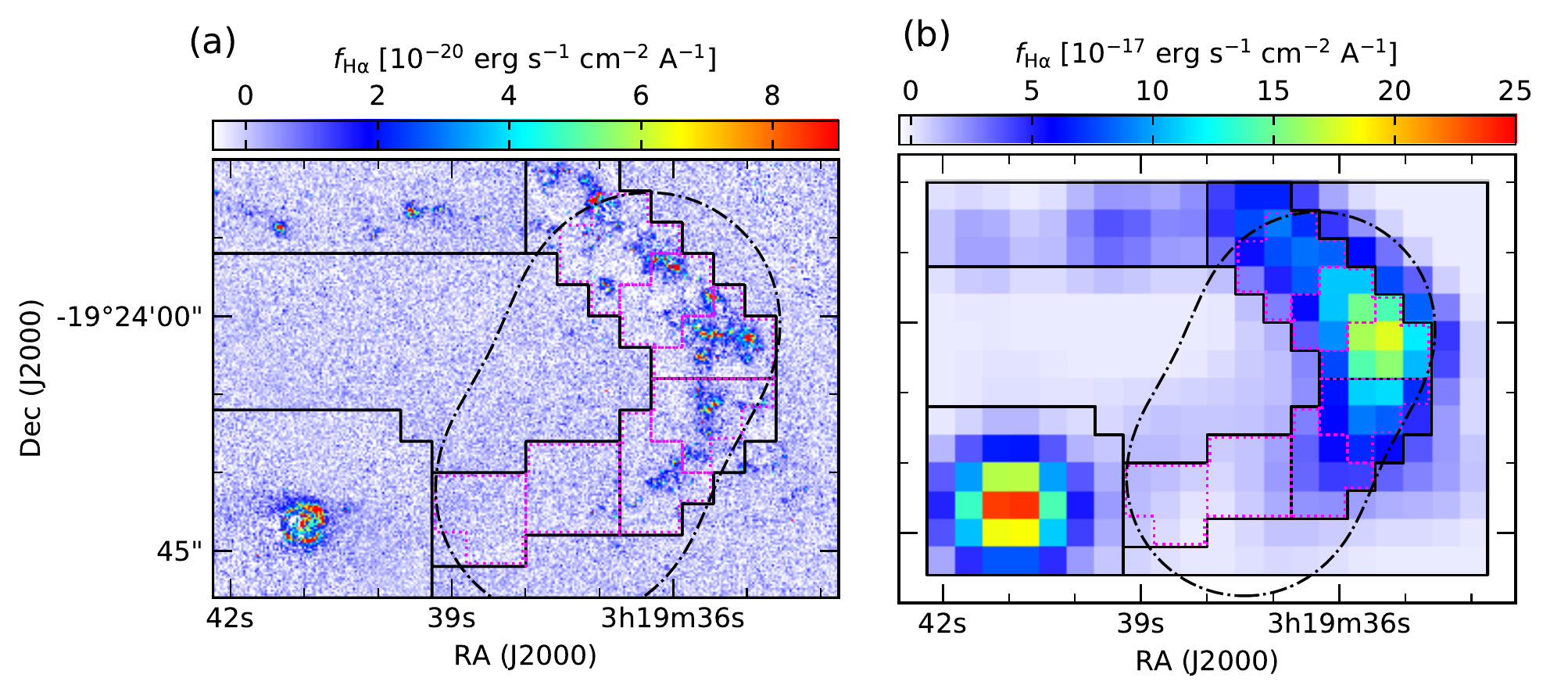}
    \caption{(a) Continuum-subtracted H$\alpha$ image made from HST images (see text). Pixel size is 0.04 arcsec.
    (b) Continuum-subtracted H$\alpha$ image convolved to the common beam size of 16.7 arcsec and regrided to 6.0 arcsec.
    All lines are the same as Fig.~\ref{fig:Result_ICO21}.}
 \label{fig:Result_Ha}
\end{figure*}

\subsubsection{H$\alpha$ emission from HST archive data} \label{sec: HST Ha}

\begin{table}
 \caption{Filter properties}
 \label{tab:HST Filter properties}
 \begin{tabular}{lccc}
\hline
Filter & \multicolumn{1}{c}{$\lambda_{\rm pivot}$$^a$} & \multicolumn{1}{c}{$W_{\rm eff}$$^b$} & \multicolumn{1}{c}{PHOTFLAM}\\
       & \multicolumn{1}{c}{(\AA)} & \multicolumn{1}{c}{(\AA)}    & \multicolumn{1}{c}{($\rm erg~cm^{-2}~s^{-1}$ \AA$^{-1}$ for 1 $\rm e^-~s^{-1}$)}  \\
  \hline
  \hline
F555W  & 5384.48     & 1099.65  &  $1.941069775 \times 10^{-19} $    \\
F658N  & 6584.06     & 74.75    &  $1.965832575 \times 10^{-18} $    \\
  \hline
\multicolumn{4}{l}{{\small$^a$  The pivot wavelength of the filter.}} \\
\multicolumn{4}{l}{{\small $^b$ The effective width defined by $\int T(\lambda) d\lambda / {\rm Max}(T(\lambda))$.}}\\
 \end{tabular}
\end{table}

\begin{figure*}
	\includegraphics[width=160mm]{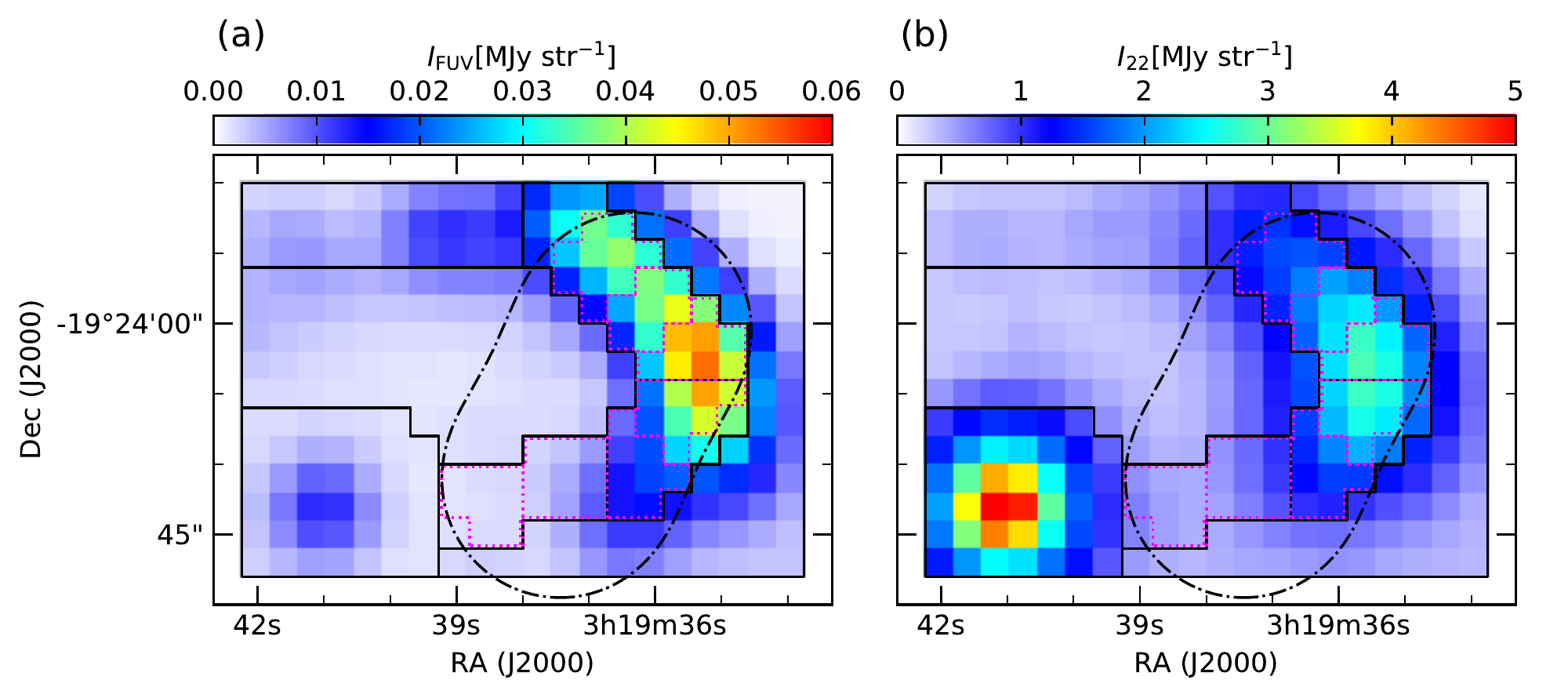}
    \caption{ (a) FUV image obtained from GALEX, but convolved to 16.7 arcsec and regrided to 6.0 arcsec.
    The black solid lines and magenta dotted lines represent definition of the environmental mask for NRO FoV  and ALMA FoV as Fig.~\ref{fig:definition_of_environments}(a) and (b), respectively. The black dash--dotted line represents the FoV observed with ALMA \citep{Maeda2020MNRAS}.
    (b) Same as panel (a), but for 22$\mu$m obtained from WISE.}
 \label{fig:Result_GALEX_WISE}
\end{figure*}

To measure SFR in NGC~1300, we made an H$\alpha$ image using the archival images observed with the ACS on the HST (proposal ID = 10342, PI = K. Noll)
which can be taken from the HLA.
The images were taken with a broadband {\it F555W} filter (close to standard $V$ band) and a narrowband H$\alpha$ filter, {\it F658N}.
The properties of these filters are summarized in Table \ref{tab:HST Filter properties}.
The observations were carried out in 2004 September and 
the total exposures time was 1360~s and 2720~s for {\it F555W} and {\it F658N}, respectively.
The fits images we used
were corrected for bias, dark current, flat-fielding, and rejection of cosmic rays
using the standard pipeline calibration for ACS.
The spatial sampling was 0.04 $\rm arcsec~\rm pixel^{-1}$.
Pixel values of the images were in $\rm electrons~s^{-1}$ and we converted this value to the flux density, unit of $\rm erg~cm^{-2}~s^{-1}$ \AA$^{-1}$, by using the calibration parameter, PHOTFLAM, listed in Table \ref{tab:HST Filter properties}.
The background was determined as a mode value, which is derived based on binning of the flux density (bin size of $1.0 \times 10^{-23} \rm erg~cm^{-2}~s^{-1}$ \AA$^{-1}$), in empty corners of the image (i.e.  region without emission from the galaxy).

The {\it F658N} image includes  H$\alpha$ emission line and  stellar continuum.
We thus subtracted the underlying stellar continuum in the {\it F658N} image adopting a method developed by \citet{Boker1999ApJS} and used by \citet{Knapen2004A&A} and \citet{Gutierrez2011AJ}.
The method is as follows: 
First, we plot the flux density of each pixel in the H$\alpha$+continuum image versus the flux density of the same pixel in the continuum image. 
Pixels which do not contain emission line delineate a straight zone in the diagram.
Pixels which contain emission line locate outside of this straight zone.
Scatter along this straight zone is originated from combination of the observation noise,
colour variations due to different stellar populations,
differential extinction effects, and so on.
The slope of this straight zone gives the constant of proportionality for the continuum subtraction.
Therefore, we can produce the continuum-subtracted H$\alpha$ image by subtracting the continuum image multiplied by this constant from the original H$\alpha$ image.
In this study,
we use the {\it F555W} image as a continuum image assuming that the flux density in the F658N image ($f_{\rm F658N}$) is proportional to the flux density in the {\it F555W} image ($f_{\rm F555W}$) in regions without H$\alpha$ emission.

Considering that the stellar population (i.e. colour) is different with environments, we determined the constant in each environment defined as Fig.~\ref{fig:definition_of_environments}(a). 
Fig.~\ref{fig:continuum_fitting_result} shows $f_{\rm F658N}$ of each pixel versus $f_{\rm F555W}$ of the same pixel in each environment. In each panel, a straight zone is seen. Then we fit the straight zone to a line as
\begin{equation}
    f_{\rm F658N} = a \times f_{\rm F555W} + b,
\end{equation}
where we added an intercept of $b$ because fitting does not work well with one parameter of $a$. 
If ordinary least squares fitting is used, the fitting will fail because the pixels containing H$\alpha$ emission, which become outliers, hinder the calculation. Therefore, we fit the straight zone using robust least-squares regression in order to minimize the influence of the outliers. 
We use Tukey's biweight to down-weight outliers:
\begin{eqnarray}
       w(d) =  \left\{
    \begin{array}{l}
         \left[ 1 - \left( \frac{d}{W} \right)^2 \right]^2 \ \ (|d| \leq W)
 \\
      0 \ \ (|d| > W),
    \end{array}
  \right.
\end{eqnarray}
where $d$ is the distance between the point and the regression line, and $W$ is a user selected value
that is most often in the range of 4-6. We applied $W = 4.0$, but changing $W$ between 4 and 6 does not influence on the fitting result. We start with an estimate of the regression line with $w(d) = 1.0$ (e.g. ordinary standard least squares). Then, we compute the Tukey's weight based on the estimated regression line. Next, we perform weighted least squares with the weight to get a new estimate of the regression line. This is repeated until $a$ and $b$ are converged. The best fit lines are shown as red solid lines in Fig.~\ref{fig:continuum_fitting_result}. The slope of the best fit line depends on the environments as we expected: The $a$ ranges from 0.6 to 1.2.

Finally, we measure the flux density of the H$\alpha$ ($f_{\rm H\alpha}$) as 
\begin{equation}
    f_{\rm H\alpha} = f_{\rm F658N} - ( a \times f_{\rm F555W} + b).
\end{equation}
Fig.~\ref{fig:Result_Ha}(a) shows the continuum-subtracted H$\alpha$ image.
Fig.~\ref{fig:Result_Ha}(b) is the image convolved to the common beam size of 16.7 arcsec and regrided to 6.0 arcsec.
In order to estimate the uncertainty of $f_{\rm H\alpha}$, we randomly extracted $10^4$ pixels from each environment and performed the fitting described above 10 times. The half of the difference when the H$\alpha$ luminosity is maximum and minimum in the 10 iterations is adopted as the uncertainty, leading typically 10~percent uncertainty of $f_{\rm H\alpha}$ in the 6.0 arcsec grid.

Because the narrowband {\it F658N} filter covers
the $\rm [N_{II}]$ doublet, $\lambda \lambda$6548, 6584, emission lines, $f_{\rm H\alpha}$ is contaminated.
Flux density ratios of $F_{{\rm [N_{II}]}_{\lambda 6548}}/F_{{\rm [N_{II}]}_{\lambda 6584}}$ 
and $F_{{\rm [N_{II}]}_{\lambda 6584}}/F_{{\rm H}\alpha}$ 
are $0.33$ and $0.30$, respectively, in the normal HII regions \citep{Osterbrock1989}. Considering the transmission corrections, we estimate the $\rm [N_{II}]$ contamination of 21~percent to $f_{\rm H\alpha}$. In the estimation of the SFR from the H$\alpha$ emission in Section~\ref{sec: conversion to SFR Ha}, this contamination is corrected by multiplying the correction factor of $C_{{\rm [N_{II}]}} = 0.79$.
It is also possible that the $f_{\rm H\alpha}$ is contaminated from the [O\textsc{iii}]$\lambda$5007, which is the strongest emission line in the broadband {\it F555W} filter. The flux density ratio of $F_{{\rm [O_{III}]}_{\lambda 5007}}/F_{{\rm H}\alpha}$ ranges from 0.1 to 0.6 in the normal H\textsc{ii} regions \citep[e.g. M51;][]{Bresolin2004ApJ}. Assuming that the $a$ is typically 1.0, we estimate that the [O\textsc{iii}] contamination is up to 3~percent. Since this value is small, we need not be worried about the [O\textsc{iii}] contamination.

\subsubsection{FUV + 22$\mu \rm m$} \label{sec: GALEX WISE}
To measure SFR, we also use GALEX FUV \citep{GildePaz2007ApJS} and WISE 22 $\mu$m \citep{Wright2010AJ} archival data as an alternative to using H$\alpha$ image.
We converted the unit of the original FUV fits image, count pixel second (CPS), into $\rm MJy~sr^{-1}$. Because the angular resolution and pixel size of the original FUV fits image are 4.3 arcsec and  1.5 arcsec, respectively, this image was convolved to the common beam size of 16.7 arcsec and regrided into 6.0 arcsec shown as Fig.~\ref{fig:Result_GALEX_WISE}(a).
For the original WISE 22$\mu$m image, we first subtracted the background level, which is determined as the mode value at the blank sky. Then, we converted the unit of digital number into $\rm MJy~sr^{-1}$.
Since the effective angular resolution of the 22$\mu$m atlas image is 16.8 arcsec  (Cutri et al. 2012)\footnote{Cutri, R. M., et al. 2012, Explanatory Supplement to the WISE All-Sky Data Release Products \url{http://wise2.ipac.caltech.edu/docs/release/allsky/expsup/}.} which is almost the same as that of our $^{12}$CO($1-0$) data obtained from the NRO 45-m, we did not convolve the 22$\mu$m image.
Fig.~\ref{fig:Result_GALEX_WISE}(b) shows the 22$\mu$m image regrided into 6.0 arcsec.

\begin{figure*}
 	\includegraphics[width=\hsize]{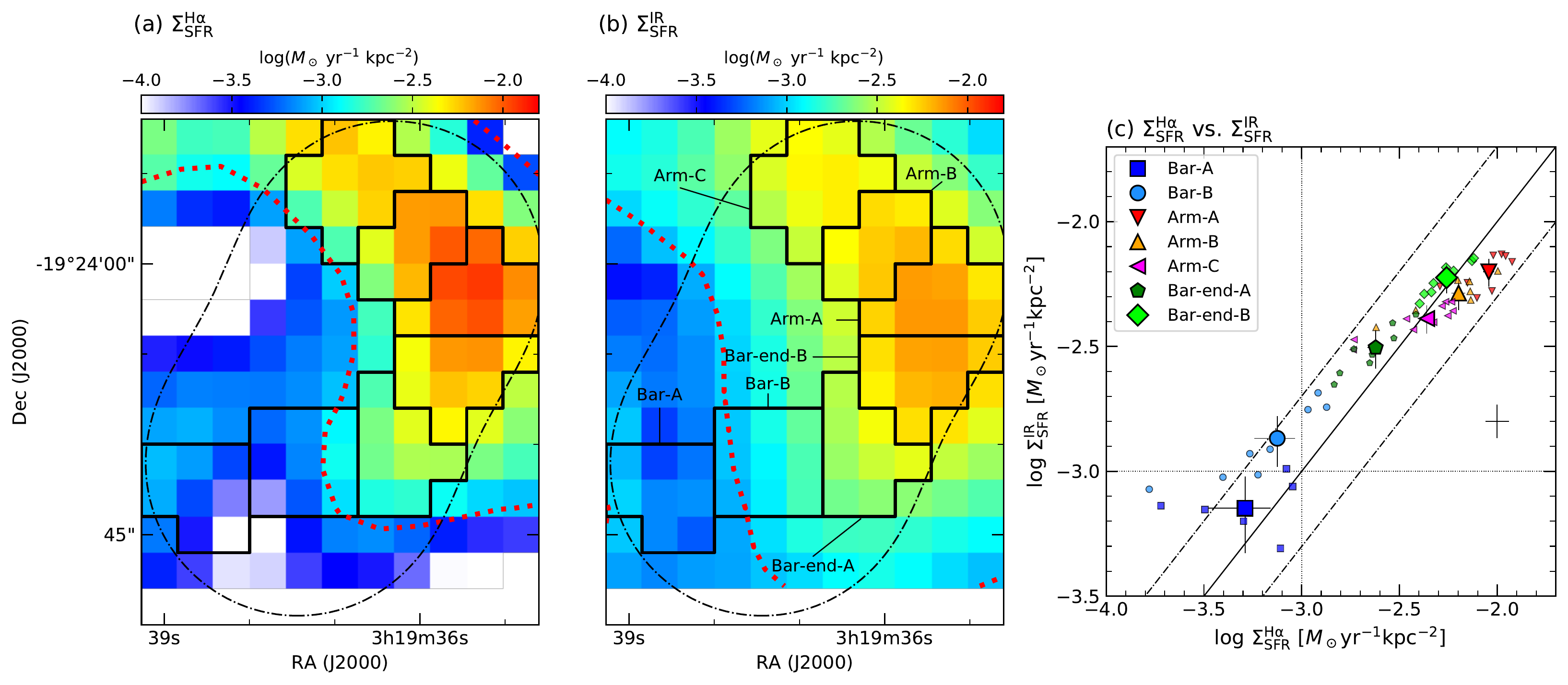}
    \caption{
    (a) Spatial distribution of surface density of SFR derived from H$\alpha$ image in ALMA FoV.
    The red dotted line represents  $10^{-3}~M_\odot~\rm yr^{-1}~kpc^{-2}$.
    Other lines are the same as Fig.~\ref{fig:Result_ICO10_ALMA}. 
    (b) Same as panel (a), but for SFR derived from FUV and 22$\mu$m images.
    (c) Comparison between surface densities of SFR derived from H$\alpha$ image and those from  FUV and 22$\mu$m images in ALMA FoV. Small and large symbols represent the value in individual 6.0 arcsec pixels and the mean value in each environment, respectively.
    The typical uncertainty for each 6.0 arcsec pixel is shown as a black cross.
    The dotted lines represent  $10^{-3}~M_\odot~\rm yr^{-1}~kpc^{-2}$.
    The solid line and dash--dotted lines represent the 1:1 correlation and factor of 2.0 differences, respectively.
    }
 \label{fig:Result_SFR_map_and_SFR_comparison}
\end{figure*}

\section{Conversion to physical parameters} \label{sec: Conversion to physical parameters}
Since the purpose of this study is investigating the differences in the $f_{\rm dif}$ with environments and the relations between $f_{\rm dif}$ and other parameters, we 
focus on the physical parameters in the environments defined for the ALMA FoV (Section~\ref{sec: NRO CO(1-0), definition of environments}) in the rest of this paper. The physical parameters in other regions including the centre are described in Appendix \ref{apx}.

\subsection{Molecular gas surface density} \label{sec: conversion to Smol}
Molecular gas surface density ($\Sigma_{\rm mol}^{\rm tot}$) is derived from the velocity-integrated intensities ($I_{\rm ^{12}CO(1-0)}^{\rm tot}$) as
\begin{equation}
    \left( \frac{\Sigma_{\rm mol}^{\rm tot}}{M_\odot \rm  pc^{-2}} \right)  =
    \left( \frac{\alpha_{\rm CO}}{M_\odot (\rm K~km~s^{-1}~pc^2)^{-1}} \right)
    \left( \frac{I_{\rm ^{12}CO(1-0)}^{\rm tot}}{\rm K~km~s^{-1}} \right)  
    \cos i,
\end{equation}
where $\alpha_{\rm CO}$ is CO-to-H$_2$ conversion factor and $\cos i$ shows the effect of the inclination of the galaxy.
Here, we adopt the standard $\alpha_{\rm CO}$ of 4.4 $M_\odot (\rm K~km~s^{-1}~pc^2)^{-1}$ including a factor of 1.36 to account for the presence of helium. 
Among the bar, arm and bar-end regions, $\Sigma_{\rm mol}$ is comparable and about $10 - 15~\rm K~km~s^{-1}$.
As described in \citet{Maeda:2018bg}, this result indicates that the molecular gases do exist in the strong bars with no clear H\textsc{ii} regions.
The range of $10 - 15~ M_\odot \rm pc^{-2}$ is consistent with those in the disc of nearby non-barred spiral galaxies \citep[e.g.][]{Bigiel2008AJ}.
Compared bar regions of other barred galaxies, the $\Sigma_{\rm mol}$ in the bar regions of NGC~1300 is lower;
Half of $\Sigma_{\rm mol}$ in NGC~1530 ( $\sim 24~M_\odot~\rm pc^{-2}$;  \citealt{ReynaudDownes1998A&A}) and NGC~5383
($ \sim 30~M_\odot~\rm pc^{-2}$;  \citealt{Maeda:2018bg}).
Less than 4 times $\Sigma_{\rm mol}$  of 
NGC~3627 ($ \sim 60~M_\odot~\rm pc^{-2}$; \citealt{Regan1999, Kuno2007PASJ}),
NGC~2903 ($ \sim 50~M_\odot~\rm pc^{-2}$; \citealt{Muraoka:2016ip})
and NGC~4303 ($ \sim 86~M_\odot~\rm pc^{-2}$; \citealt{Yajima:2019do}).
Here the $\alpha_{\rm CO}$ is unified to $4.4~M_\odot~(\rm K~km~s^{-1}~pc^2)^{-1}$.

\citet{Sorai2012PASJ} suggest that $\alpha_{\rm CO}$ in the bar regions may be $0.5-0.8$ times smaller than that in the arm regions in Maffei 2 using large velocity gradient analysis. 
\citet{MorokumaMatusi2015PASJ} suggested the existence of non-optically-thick components of $^{12}$CO($1-0$) in the bar regions may make the $\alpha_{\rm CO}$ by a factor of a few smaller than in the arm regions in NGC~3627. 
A similar result is reported by \citet{Watanabe2011}.
Therefore, 
there is a possibility that the $\Sigma_{\rm mol}$ in the bar regions of NGC~1300 may be overestimated by a factor of a few. However, this issue is beyond our scope and remains as future work.

\begin{table*}
 \caption{Mean values of physical parameters for ALMA FoV.}
 \label{tab:Stacking Analysis ALMAFoV}
 \begin{tabular}{llccccccc}
  \hline
& & Bar-A & Bar-B & Arm-A & Arm-B & Arm-C & Bar-end-A & Bar-end-B\\
\hline
$\Sigma_{\rm SFR}^{\rm H\alpha}$ &($10^{-3}~M_\odot \rm~yr^{-1}~kpc^{-2}$)  &$0.5\pm0.2$&$0.8\pm0.2$&$9.1\pm0.6$&$6.3\pm0.5$&$4.4\pm0.4$&$2.4\pm0.2$&$5.5\pm0.3$ \\
$\Sigma_{\rm SFR}^{\rm IR}$ &($10^{-3}~M_\odot \rm~yr^{-1}~kpc^{-2}$) &$0.7\pm0.2$&$1.4\pm0.3$&$6.3	\pm0.8$&$5.2\pm0.8$&$4.1\pm0.5$&$3.1\pm0.6$&$6.0\pm0.8$ \\
$I_{\rm ^{12}CO(1-0)}^{\rm tot}$ &  ($\rm K~km~s^{-1}$) &$4.4\pm0.5$&$4.1\pm0.3$&$4.6\pm0.4$&$5.3\pm0.4$&$3.0\pm0.3$&$4.9\pm0.3$&$4.7\pm0.3$\\
$I_{\rm ^{12}CO(1-0)}^{\rm ALMA}$ & ($\rm K~km~s^{-1}$) &$0.40\pm0.02$&$1.05\pm0.01$&$2.08\pm0.02$&$1.89\pm0.03$&$1.32\pm0.03$&$2.29\pm0.02$&$3.40\pm0.02$ \\
$I_{\rm ^{12}CO(2-1)}^{\rm tot}$ & ($\rm K~km~s^{-1}$) &$0.7\pm0.1$ &$1.4\pm0.1$ &$2.3\pm0.2$ &$2.2\pm0.2$ &$1.7\pm0.2$ &$2.4\pm0.2$ &$3.4\pm0.3$\\
$\Sigma_{\rm mol}^{\rm tot}$ & ($M_\odot \rm~pc^{-2}$) &$12.3\pm1.3$&$11.4\pm0.9$&$12.9\pm1.1$&$15.0\pm1.1$&$8.6\pm0.8$&$13.9\pm0.9$&$13.3\pm1.0$\\
$f_{\rm dif}$ & &$0.91\pm0.10$&$0.74\pm0.07$&$0.55\pm0.05$&$0.65\pm0.06$&$0.57\pm0.06$&$0.53\pm0.05$&$0.28\pm0.02$ \\
${\rm SFE}^{\rm H\alpha}$ &($\rm Gyr^{-1}$) &$0.06\pm0.02$&$0.09\pm0.02$&$0.96\pm0.10$&$0.58\pm0.06$&$0.69\pm0.09$&$0.23\pm0.03$&$0.57\pm0.05$\\
${\rm SFE}^{\rm IR}$ &($\rm Gyr^{-1}$) &$0.08\pm0.03$&$0.16\pm0.04$&$0.66\pm0.10$&$0.47\pm0.08$&$0.65\pm0.10$&$0.31\pm0.06$&$0.61\pm0.09$\\
$R_{21/10}$ & &$0.17\pm0.02$&$0.34\pm0.04$&$0.51\pm0.07$&$0.41\pm0.05$&$0.55\pm0.07$&$0.49\pm0.06$&$0.72\pm0.09$\\
$R_{\rm 13/12}$ & &$<0.09$&$<0.09$&$<0.12$&$<0.07$&$<0.25$&$<0.08$&$<0.08$\\
\hline
 \end{tabular}
\end{table*}

\subsection{Star formation rate}\label{sec: SFR}

\subsubsection{SFR from H$\alpha$ } \label{sec: conversion to SFR Ha}
We calculated the surface density of SFR from the continuum-subtracted H$\alpha$ image (Fig.~\ref{fig:Result_Ha}(b)) by
\begin{equation}
    \left( \frac{\Sigma_{\rm SFR}^{\rm H\alpha}}{M_\odot \rm~yr^{-1}~kpc^{-2}} \right)
    =
    5.3 \times 10^{-42}
    \left( \frac{L_{\rm H\alpha}}{\rm erg~s^{-1}} \right)
    \frac{\cos i}{S},
\end{equation}
where $5.3 \times 10^{-42}$ is the conversion factor from H$\alpha$ luminosity ($L_{\rm H\alpha}$) to the SFR
obtained by \citet{Calzetti2007ApJ}, $i$ is the inclination of NGC~1300 and $S$ is the covered area for each pixel of $600~{\rm pc} \times 600~{\rm pc}$. The $L_{\rm H\alpha}$ is derived as
\begin{equation}
L_{\rm H\alpha}=
4 \pi 
f_{\rm H\alpha}
W_{\rm eff}
D^2 
C_{\rm [NII]}
10^{0.4A_{\rm V}},
\end{equation}
where $f_{\rm H\alpha}$ is the H$\alpha$ flux density derived from the continuum-subtracted H$\alpha$ image,
$W_{\rm eff}$ is the effective width of the {\it F658N} filter listed in Table \ref{tab:HST Filter properties},
$D$ is the distance to NGC~1300,
$C_{{\rm [N_{II}]}}$ is the correction factor of 0.79
which is needed to remove the $\rm [N_{II}]$ emission described in Section~\ref{sec: HST Ha}, and $A_{\rm V}$ is a correction for the dust extinction.
The dust extinction can vary in star-forming regions,
and it is difficult to determine the optical depth.
Here, we assumed an extinction of  $A_{\rm V} = 1.0$ mag, 
which is a typical value for integrated galaxy disks obtained by
\citet{Leroy2012AJ}. They investigate SFRs in 30 disc galaxies from H$\alpha$, IR, and FUV images, and found that H$\alpha$ emission contributes $\sim 40$ percent of the total SFR over the range, $\Sigma_{\rm SFR} = 10^{-1} - 10^{-3}~M_\odot~\rm yr^{-1}~kpc^{-2}$, corresponding to $\sim 1.0$ mag of extinction.
Fig.~\ref{fig:Result_SFR_map_and_SFR_comparison}(a) shows the spatial distribution of $\Sigma_{\rm SFR}^{\rm H\alpha}$.
While the $\Sigma_{\rm SFR}^{\rm H\alpha}$ is $10^{-2.7}-10^{-1.9}~M_\odot~\rm yr^{-1}~kpc^{-2}$ in arm and bar-end regions, $\Sigma_{\rm SFR}^{\rm H\alpha}$ is mostly lower than $10^{-3}~M_\odot~\rm yr^{-1}~kpc^{-2}$ in bar (and inter-arm) regions.
The uncertainty of $\Sigma_{\rm SFR}^{\rm H\alpha}$ is typically 10~percent propagated from the determination of $f_{\rm H\alpha}$ described in Section~\ref{sec: HST Ha}.

\subsubsection{SFR from FUV and 22$\mu$m} \label{sec: conversion to SFR IR}

The surface density of SFR can be calculated from a linear combination of GALEX FUV and WISE 22$\mu$m intensities by \citet{Casasola2017A&A} as
\begin{eqnarray}
    \left( \frac{\Sigma_{\rm SFR}^{\rm IR}}{M_\odot \rm~yr^{-1}~kpc^{-2}} \right)
    = \nonumber \\
    \left[ 8.1 \times 10^{-2}
    \left( \frac{I_{\rm FUV}}{ \rm~MJy~str^{-1}} \right)
    + 3.2 \times 10^{-3}
    \left( \frac{I_{\rm 22\mu m}}{ \rm~MJy~str^{-1}} \right) \right] \cos i \label{eq: SFR IR},
\end{eqnarray}
where $I_{\rm FUV}$ and $I_{\rm 22\mu m}$ 
are the FUV and 22 $\mu$m intensities, respectively.
Note that the equation (\ref{eq: SFR IR}) is the relation presented by \citet{Leroy2008AJ.136.2782L} by replacing 24 $\mu$m intensity with the 22 $\mu$m one. 

To estimate the SFR from IR emission, contamination from the older stellar population need to be considered. Qualitatively, the dust emission from old stars represents a second-order correction, but still important in nearby disc galaxy \citep{Law2011ApJ, Leroy2012AJ}.
We remove the 22 $\mu$m emission not associated with recent SF using following equation:
\begin{equation}
    \left( \frac{I_{\rm 22\mu m}^{\rm old}}{\rm MJy~str^{-1}} \right) = 1.2 \times 10^{-2} \left( \frac{\Sigma_{\rm gas}}{M_\odot~\rm pc^{-2}} \right),
\end{equation}
where $\Sigma_{\rm gas}$ is the gas surface density combines H\textsc{i} and H$_2$. This equation is presented by \citet{Leroy2012AJ} and we replaced 24 $\mu$m intensity with the 22 $\mu$m one. Comparing the H\textsc{i} image \citep{England1989a} and H$_2$ map we obtained, we assume $\Sigma_{\rm H_2}/(\Sigma_{\rm H_I}+\Sigma_{\rm H_2}) = 0.7$. 
The derived $I_{\rm 22\mu m}^{\rm old}$ is typically $0.1\sim0.3~{\rm MJy~str^{-1}}$ and contribution from IR emissions originated from old stars to $\Sigma_{\rm SFR}^{\rm IR}$ is $5-10$ percent in {\it arm-A}, {\it arm-B}, {\it arm-C} and {\it bar-end-B} and $15 -25$ percent in {\it bar-end-A}, {\it bar-A}, and {\it bar-B}. Old stars may still produce some FUV emission, which also contaminates SFR estimates.  However, such contamination appears to be only 3 percent of the total SFR where $\Sigma_{\rm SFR} > 10^{-3}~M_\odot \rm~yr^{-1}~kpc^{-2}$ \citep{Leroy2012AJ}. Thus, we neglected this contamination in this study.

\begin{figure*}
	\includegraphics[width=\hsize]{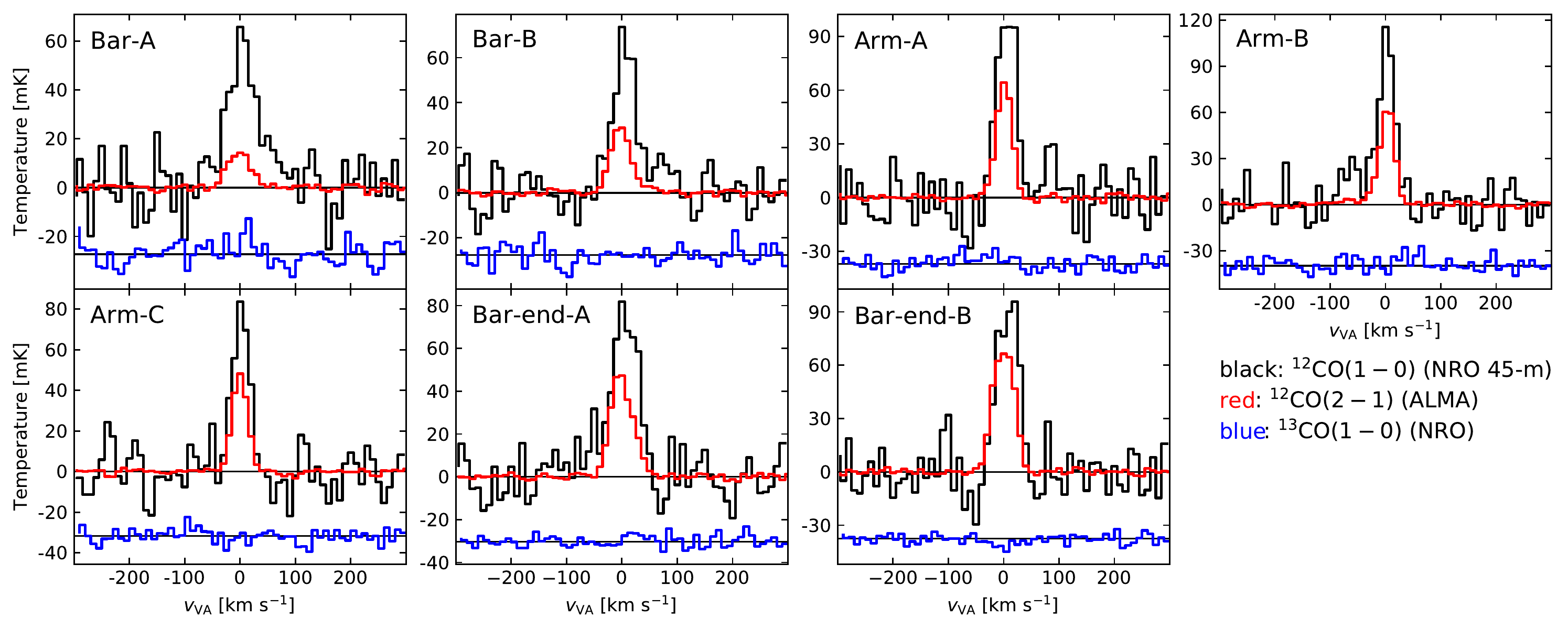}
    \caption{Stacked CO spectra for each environment in NGC~1300. The black, red, and blue lines indicate $^{12}$CO($1-0$) emission obtained from the NRO 45-m, $^{12}$CO($2-1$) emission obtained from ALMA, and  $^{13}$CO($1-0$) emission obtained from the NRO 45-m, respectively.
    The channel width is $10.0~\rm km~s^{-1}$
    For visualization purpose, base line of $^{13}$CO($1-0$) spectra is shifted.}
 \label{fig:Result_stacking_analysis}
\end{figure*}

Fig.~\ref{fig:Result_SFR_map_and_SFR_comparison}(b) shows the spatial distribution of $\Sigma_{\rm SFR}^{\rm IR}$.
$\Sigma_{\rm SFR}^{\rm IR}$ is $10^{-2.5} - 10^{-2.1}~M_\odot~\rm yr^{-1}~kpc^{-2}$ in arm and bar-end regions and $<10^{-2.7}~M_\odot~\rm yr^{-1}~kpc^{-2}$ in bar regions.
The uncertainty of $\Sigma_{\rm SFR}^{\rm IR}$ is typically
16~percent, which includes the uncertainty of contamination from the older stellar population and calibration uncertainties of $I_{\rm FUV}$ and $I_{\rm 22\mu m}$; The FUV zero-point calibration is estimated to be 0.15 mag, which yields the uncertainty of $\sim 15$ percent \citep{GildePaz2007ApJS}.
The photometric accuracy of the WISE calibrators is 5.7percent for the 22 $\rm \mu m$ band \citep{Jarrett2011ApJ}.

For our calibration of $\Sigma_{\rm SFR}^{\rm H\alpha}$ and $\Sigma_{\rm SFR}^{\rm IR}$, we adopt the initial mass function (IMF) from \citet{Calzetti2007ApJ}.
To convert the $\Sigma_{\rm SFR}$ to those shown by \citet{Kennicutt1998ARA&A} adopted the truncated \citet{Salpeter1955ApJ} IMF, a factor of 1.59 should be multiplied.

\begin{figure*}
	\includegraphics[width=\hsize]{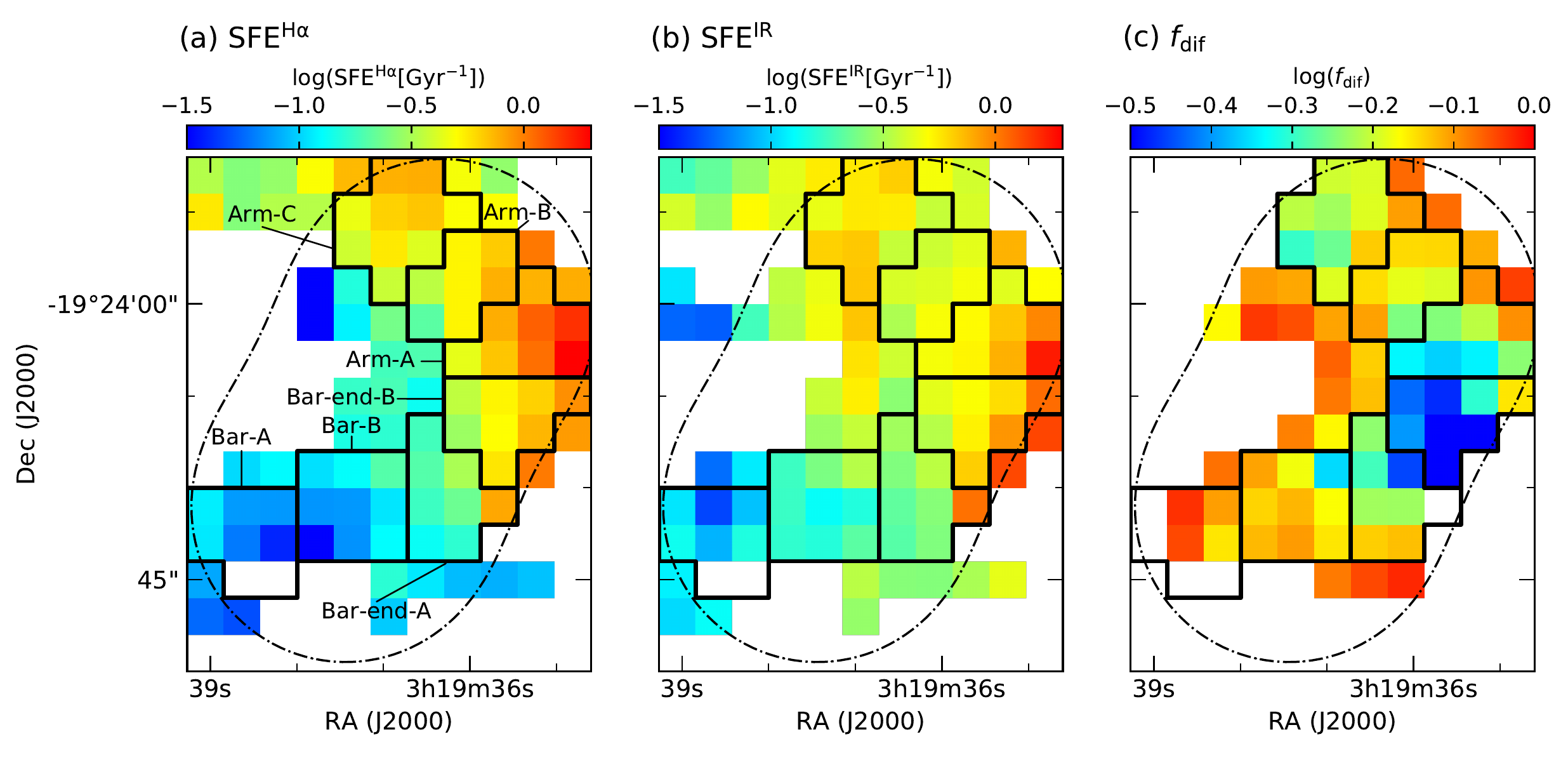}
    \caption{(a) Spatial distribution of the SFE obtained from H$\alpha$ image.
    All lines are the same as Fig.~\ref{fig:Result_ICO10_ALMA}. 
    (b) Same as panel (a), but for the SFE obtained from FUV and 22$\mu$m image.
    (c) Same as panel (a), but for the diffuse molecular gas fraction.
    }
 \label{fig:Result_fre_SFE_map_in_ALMAFoV}
\end{figure*}

\subsubsection{SFR Comparison} \label{sec: SFR Comparison}

In order to check the accuracy of the $\Sigma_{\rm SFR}$ measurements, we compared $\Sigma_{\rm SFR}^{\rm H\alpha}$ and $\Sigma_{\rm SFR}^{\rm IR}$ in Fig.~\ref{fig:Result_SFR_map_and_SFR_comparison}(c).
The small symbols represent individual 6.0 arcsec pixels.
The mean value of the $\Sigma_{\rm SFR}$ in each environment is shown as a large symbol and listed in Table \ref{tab:Stacking Analysis ALMAFoV}. In {\it arm-A}, {\it arm-B}, and {\it arm-C} regions, $\Sigma_{\rm SFR}^{\rm H\alpha}$ is systematically larger than $\Sigma_{\rm SFR}^{\rm IR}$ by a factor up to $1.4$.
While in {\it bar-end-A} and {\it bar-end-B} regions, $\Sigma_{\rm SFR}^{\rm H\alpha}$ is systematically smaller than $\Sigma_{\rm SFR}^{\rm IR}$ by a factor of $0.8-0.9$. The cause for these differences is unknown, but there are some possibilities: differences of dust extinction among environments, uncertainties in estimations of the 22$\mu$m emission not associated with recent SF, and so on. In any case, the magnitude of systematic differences rarely exceeds a factor of two (0.3 dex) above $\Sigma_{\rm SFR} = 10^{-3.0}~M_\odot~\rm yr^{-1}~kpc^{-2}$.

Below $\Sigma_{\rm SFR}^{\rm H\alpha} = 10^{-3.0}~M_\odot~\rm yr^{-1}~kpc^{-2}$, or in {\it bar-A}, {\it bar-B}, the both values differ by more than a factor of two in some pixels; $\Sigma_{\rm SFR}^{\rm IR}$ is larger than $\Sigma_{\rm SFR}^{\rm H\alpha}$ by a factor up to $5.0$.
Such large difference below $\Sigma_{\rm SFR} = 10^{-3.0}~M_\odot~\rm yr^{-1}~kpc^{-2}$ is also reported by \citet{Leroy2012AJ}.
They claim the cause for the difference is that the measured H$\alpha$ surface brightness of $\Sigma_{\rm SFR} \leq 10^{-3.0}~M_\odot~\rm yr^{-1}~kpc^{-2}$ is most likely not originated from massive SF and $\Sigma_{\rm SFR}^{\rm IR}$ strongly depends on the method to remove IR cirrus.
Because of the absence of H\textsc{ii} regions, the
$\Sigma_{\rm SFR}^{\rm H\alpha}$  in {\it bar-A} seems to be dominated by the noise and the smear of the H$\alpha$ emission in the centre region by the convolution. 
$\Sigma_{\rm SFR}^{\rm H\alpha}$ can be underestimated because ionising photons can escape HII regions, but some extragalactic studies suggest that this effect might be small when averaging at extragalactic scales \citep[e.g.][]{Murphy2011ApJ,Querejeta:2019jl}.
In summary, the accuracy of the $\Sigma_{\rm SFR}$ above $10^{-3.0}~M_\odot~\rm yr^{-1}~kpc^{-2}$ seems to be high, but below $10^{-3.0}~M_\odot~\rm yr^{-1}~kpc^{-2}$, the measured $\Sigma_{\rm SFR}$ may not indicate the exact amount of massive SF and should be considered to be an upper limit.

\subsection{Stacking Analysis} \label{sec: Result Stacking Analysis}
Since the rms noise level of $^{12}$CO($1-0$) spectra obtained from the NRO 45-m is high, 
spatial distribution of $f_{\rm dif}$ and $R_{21/10}$ seem to be slightly noisy. In this section,
in order to improve the S/N of the $^{12}$CO($1-0$) emission, we performed the stacking analysis of CO spectra with velocity axis alignment devised by \citet{Schruba2011AJ,Schruba2012AJ}. 
This method is commonly adopted for the CO spectra in nearby galaxies \citep[e.g.][]{MorokumaMatusi2015PASJ,Muraoka:2016ip,Yajima:2019do}.
The procedure is as follow:
First, the velocity field of NGC~1300 is estimated from $^{12}$CO($2-1$) data obtained from ALMA. We calculated the intensity-weighted mean velocity ($\bar{v}_{\rm CO(2-1)}$) of each pixel. Then, the $^{12}$CO($1-0$) spectra are shifted along the velocity axis based on the $\bar{v}_{\rm CO(2-1)}$. Finally, we made an average $^{12}$CO($1-0$) spectrum in each environment. For $^{13}$CO($1-0$) spectra and $^{12}$CO($1-0$) spectra obtained from ALMA, we adopted this method.
Fig.~\ref{fig:Result_stacking_analysis} shows the stacked CO spectra. 
Black line represents the stacked $^{12}$CO($1-0$) spectra. The rms noise level is improved ($8 \sim 14$ mK at 10.0 $\rm km~s^{-1}$).
Red and blue line represent the stacked $^{12}$CO($2-1$) and $^{13}$CO($1-0$) spectra, respectively. 
In Table \ref{tab:Stacking Analysis ALMAFoV}, we listed the velocity integrated intensity of each stacked spectrum.
Although the rms noise level of $^{13}$CO($1-0$) spectra is improved ($5 \sim 8$ mK at 10.0 $\rm km~s^{-1}$), we did not detect the significant $^{13}$CO($1-0$) emission line in any environments.

For $^{12}$CO($1-0$) data obtained from ALMA, the mean value of the pixels in each environment is adopted as a stacked value.
This is because significant negative structures ("bowls") in the data and this structures may influence on the stacking analysis of the spectra with velocity axis alignment. These negative bowls are expected to be present in interferometric data
because our data are missing the shortest $uv$ spacings.

\begin{figure*}
	\includegraphics[width=150mm]{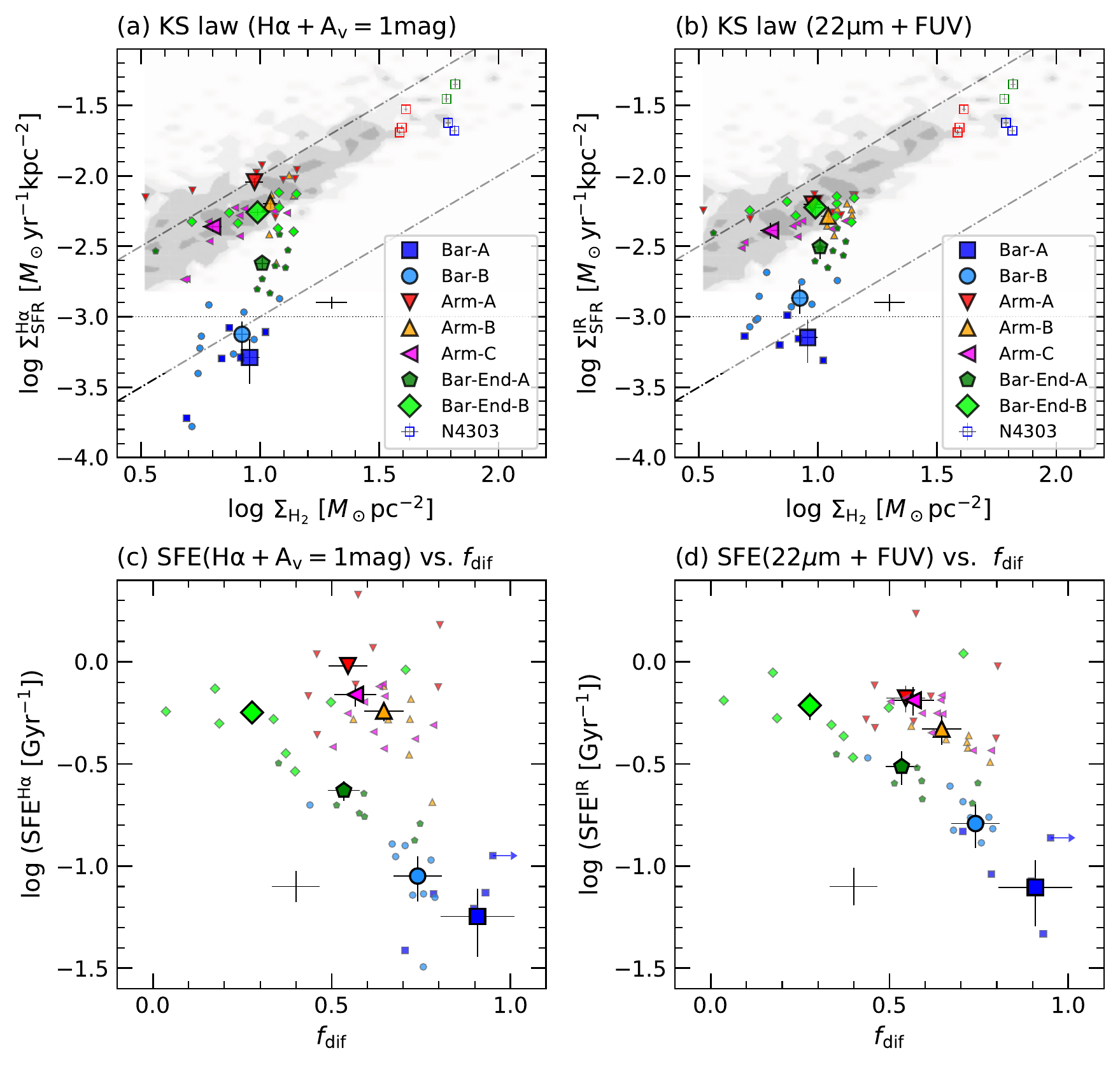}
    \caption{(a) Molecular Kennicutt-Schmidt relation of NGC~1300. SFR is derived from the H$\alpha$ emission. 
    The small and large symbols represent the value in individual 6.0 arcsec pixels and the mean value in each environment, respectively.
    The typical uncertainty for each 6.0 arcsec pixel is shown as a black cross.
    The grey scale represents the result by \citet{Bigiel2008AJ}. 
    The open squares represents the result in NGC~4303 by \citet{Yajima:2019do} (blue: bar, red: arm, and green: bar-end). 
    The grey dash--dotted lines show the constant SFE of $1.0$ and $0.1~\rm Gyr^{-1}$.
    The horizontal dotted line represents $\Sigma_{\rm SFR} = 10^{-3}~M_\odot~\rm yr^{-1}~kpc^{-2}$.
    (b) Same as panel (a), but SFR is derived from FUV and IR.
    (c) Correlation of diffuse molecular gas fraction, $f_{\rm dif}$ with SFE. SFR is derived from the H$\alpha$ emission.  
    (d) Same as panel (c), but SFR is derived from FUV and IR.}
 \label{fig:Result_ALMAFoV_KS_fre_SFE}
\end{figure*}

\section{Star formation efficiency and recovery fraction} \label{sec: Star formation efficiency and recovery fraction}
\subsection{Star formation efficiency} \label{sec: Result SFE}
As described in Section~\ref{sec: Intro}, SF activity is quite different among the environments in NGC~1300. In this section, we examine the difference  quantitatively through the SFE.
SFE is calculated using the surface density of SFR and that of molecular hydrogen ($\Sigma_{\rm H_2}^{\rm tot}$) as follows: 
\begin{equation}
    \left( \frac{\rm SFE}{\rm Gyr^{-1}} \right) = 
    10^3
    \left( \frac{\Sigma_{\rm SFR}}{M_\odot~\rm yr^{-1}~kpc^{-2}} \right)
    \left( \frac{\Sigma_{\rm H_2}^{\rm tot}}{M_\odot~\rm pc^{-2}} \right)^{-1},
\end{equation}
where $\Sigma_{\rm H_2}^{\rm tot}$ is $\Sigma_{\rm mol}^{\rm tot}/1.36$ to remove the helium.
The SFE obtained from $\Sigma_{\rm SFR}^{\rm H\alpha}$ and $\Sigma_{\rm SFR}^{\rm IR}$, which are denoted as $\rm {SFE}^{\rm H\alpha}$ and $\rm {SFE}^{\rm IR}$, respectively. 
Fig.~\ref{fig:Result_fre_SFE_map_in_ALMAFoV}(a) and (b) show the spatial distribution of $\rm {SFE}^{\rm H\alpha}$ and $\rm {SFE}^{\rm IR}$, respectively.
We display the pixels where both the SFR tracers and $^{12}$CO($1-0$) emission observed with the NRO 45-m were significantly detected. The mean values obtained from the stacked $^{12}$CO($1-0$) spectra and mean $\Sigma_{\rm SFR}$ in each environment are listed in Table \ref{tab:Stacking Analysis ALMAFoV}.
The difference of the SFE with the environments is clearly seen: In bar regions, although the uncertainty of $\Sigma_{\rm SFR}$ is large (Section \ref{sec: SFR Comparison}), SFE is very low $(0.06-0.16)~\rm Gyr^{-1}$
In {\it bar-end-A} near the bar region, SFE is low $(0.23-0.31)~\rm Gyr^{-1}$ in comparison to other arm and bar-end regions. 
The SFE is the highest in {\it arm-A} $(0.66-0.96)~\rm Gyr^{-1}$, followed 
in {\it bar-end-B}, {\it arm-B}, {\it arm-C} $(0.47-0.69)~\rm Gyr^{-1}$.
In summary, SFEs in the bar regions are more than $\sim 5 - 10$ times lower than those in the arm regions. In spatial distribution of SFE, gradient in SFE from leading to trailing side is also seen in the arm and bar-end regions. This gradient is originated from the offsets between SFR tracers (H$\alpha$) and CO emission, which are often seen in spiral galaxies \citep[e.g.][]{Schinnerer:2013jy,Schinnerer2017ApJ836,Kreckel:2018ep}. 

The differences of the SFE with environments are also clearly seen in the molecular redKennicutt-Schmidt diagram, or $\Sigma_{\rm H_2}^{\rm tot}$ vs. $\Sigma_{\rm SFR}$, as shown in Fig.~\ref{fig:Result_ALMAFoV_KS_fre_SFE}(a) and (b).
Large symbols show the mean values in the environments, which are listed in Table \ref{tab:Stacking Analysis ALMAFoV}.
The grey scale represents the result by \citet{Bigiel2008AJ}, which derived K-S relations at an angular resolution of 750~pc in a sample of 18 nearby non-barred spiral galaxies.
SFRs were measured from a combination of FUV and 24$\mu$m IR fluxes and H$_2$ gases were measured from $^{12}$CO($2-1$) assuming the line ratio of $R_{21/10} = 0.8$. Open squares represent the K-S relation in NGC~4303 at an angular resolution of 1.4 kpc by \citet{Yajima:2019do}. They also used FUV and 24$\mu$m IR fluxes, but measured H$_2$ gases from $^{12}$CO($1-0$).
Black dash--dotted lines show the constant SFE of $1.0$ and $0.10~\rm Gyr^{-1}$. 

In Fig.~\ref{fig:Result_ALMAFoV_KS_fre_SFE}(a) and (b), it is clearly shown that the SFEs in the bar regions are much lower than those in the arm regions despite the  $\Sigma_{\rm H_2}$ is comparable. This result supports that the SF in the bar regions is suppressed compared to the arm regions. We also find that the SFE in {\it bar-end-A} is slightly suppressed compared to the {\it arm-A}, {\it arm-B}, {\it arm-C} and {\it bar-end-B} where the SFEs are the same as the result by \citet{Bigiel2008AJ}.
Although the $\Sigma_{\rm H_2}^{\rm tot}$ in NGC~4303 is $7 - 8$ times higher than those in NGC~1300, the SFEs in arm and bar-end regions in NGC~4303 are comparable to those in the arm regions in NGC~1300.
As seen in these panels, in intermediate-type barred galaxies like NGC~4303, the SFE in the bar regions is about $1.5 - 2.0$ times smaller than that in the arm region   \citep[see also][]{Momose2010ApJ,Hirota:2014bt}.
Our result indicates that the SF in the bar regions of the strongly barred galaxies
is further suppressed than in the intermediate-type barred galaxies.

As described in Section~\ref{sec: conversion to Smol},
the CO-to-H$_2$ conversion factor may be about 2 times smaller than $\alpha_{\rm CO} =  4.4~M_\odot~(\rm K~km~s^{-1}~pc^2)^{-1}$ in the bar regions.
If so, SFEs in the bar regions are much lower than those in the arm regions.
Therefore, we conclude that the SF activity is remarkably suppressed in the bar regions of strongly barred galaxy, NGC~1300.

\subsection{Diffuse molecular gas fraction} \label{sec: Diffuse gas fraction}

We measure the diffuse molecular gas fraction, $f_{\rm dif}$,  in NGC~1300 as 
\begin{equation}
    f_{\rm dif} = 1 - f_{\rm re} = 1 -  \frac{I_{\rm ^{12}CO(1-0)}^{\rm ALMA}}{I_{\rm ^{12}CO(1-0)}^{\rm tot}},
\end{equation}
where $f_{\rm re}$ is the recovery fraction, i.e. the ratio of CO flux obtained from interferometer of ALMA 12-m array to total CO flux obtained from the NRO 45-m.
As shown in Fig.~\ref{fig:recovery_fraction_simulation}, a molecular gas structure homogeneously extended to over $\sim700$~pc is mostly resolved out in our  $^{12}$CO($1-0$) data obtained from ALMA.
Therefore, the $f_{\rm dif}$ ($f_{\rm re}$) indicates the ratio of molecular gas on a scale of 700 pc or more (less).
Fig.~\ref{fig:Result_fre_SFE_map_in_ALMAFoV}(c) shows the spatial distribution of the $f_{\rm dif}$.
We display the pixels where both the $I_{\rm ^{12} CO(1-0)}^{\rm tot}$ and $I_{\rm ^{12}CO(1-0)}^{\rm ALMA}$ were significantly detected. 
The mean values of $f_{\rm dif}$ obtained from the stacked $^{12}$CO($1-0$) spectra and mean $I^{\rm ALMA}_{\rm ^{12}CO(1-0)}$ in each environment are listed in Table \ref{tab:Stacking Analysis ALMAFoV}.
The difference of the $f_{\rm dif}$ with the environments is seen,
and we find that  $f_{\rm dif}$ in the bar region is large. Thus significant fraction of the molecular gas is diffuse as compared with arm and bar-end regions: In {\it bar-A} and {\it bar-B}, the $f_{\rm dif}$ is the highest ($0.74 - 0.91$), which indicates most of the molecular gas in the bar region exists as diffuse gas with FWHM $\gtrapprox 700~\rm pc$. In {\it bar-end-A}, {\it arm-A}, {\it arm-B} and {\it arm-C}, the $f_{\rm dif}$ ranges $0.53-0.65$.
In {\it bar-end-A}, the $f_{\rm dif}$ is the lowest ($0.28$), which implies that the compact molecular gas components like GMCs are concentrated.

\subsection{SFE vs. diffuse molecular gas fraction} \label{sec: Result SFE vs. Recovery fraction}

Does the presence of a large amount of diffuse molecular gases really make the SFE low in the bar regions? If this scenario is true, the SFE is expected to decrease as $f_{\rm dif}$ increases. Thus we investigate the relation between SFE and $f_{\rm dif}$ as shown in Fig.~\ref{fig:Result_ALMAFoV_KS_fre_SFE}(c) and (d).
Large symbols show the mean values in the environments, which are listed in Table \ref{tab:Stacking Analysis ALMAFoV}.
We find that 
the SFE decreases with increasing $f_{\rm dif}$ for the large symbols: In the bar regions, $f_{\rm dif}$ and ${\rm SFE}$ are $0.74 -0.91$ and  $(0.06-0.16)~\rm Gyr^{-1}$, respectively, and $f_{\rm dif}$ and ${\rm SFE}$ are higher and lower than those in the arm and bar-end regions ($f_{\rm dif} = 0.28 -0.65$ and ${\rm SFE} = 0.23-0.96~\rm Gyr^{-1}$).
A tighter relationship is seen when using the FUV and 22$\mu$m as an SFR tracer rather than using H$\alpha$ emission.
This decreasing trend would support the idea that a large amount of diffuse molecular gases makes the SFE low.
In bar regions, a significant diffuse molecular component would be not directly related to the SF activity.

The arm regions show $\sim 3$ ($\sim 2$) times higher $\rm SFE^{\rm H\alpha}$ ($\rm SFE^{\rm IR}$) at a fixed $f_{\rm dif}$ as compared with the bar-end regions.  This suggests that the difference of SFE can not be explained only by the difference of $f_{\rm dif}$, implying the existence of other mechanisms which control the SFE.
This difference might reflect a difference in the process of feedback by massive stars. In the bar-end regions, because the distance between the GMCs is smaller than in the arm regions (see Fig. 12 in \citealt{Fujimoto2020MNRAS}), the stellar feedback would work well, and subsequent star formation may be unlikely to occur.
It is interesting to investigate the variations in the effect of stellar feedback with environments, but it is beyond our scope in this paper and remains as a future study.

\begin{figure}
	\includegraphics[width=90mm]{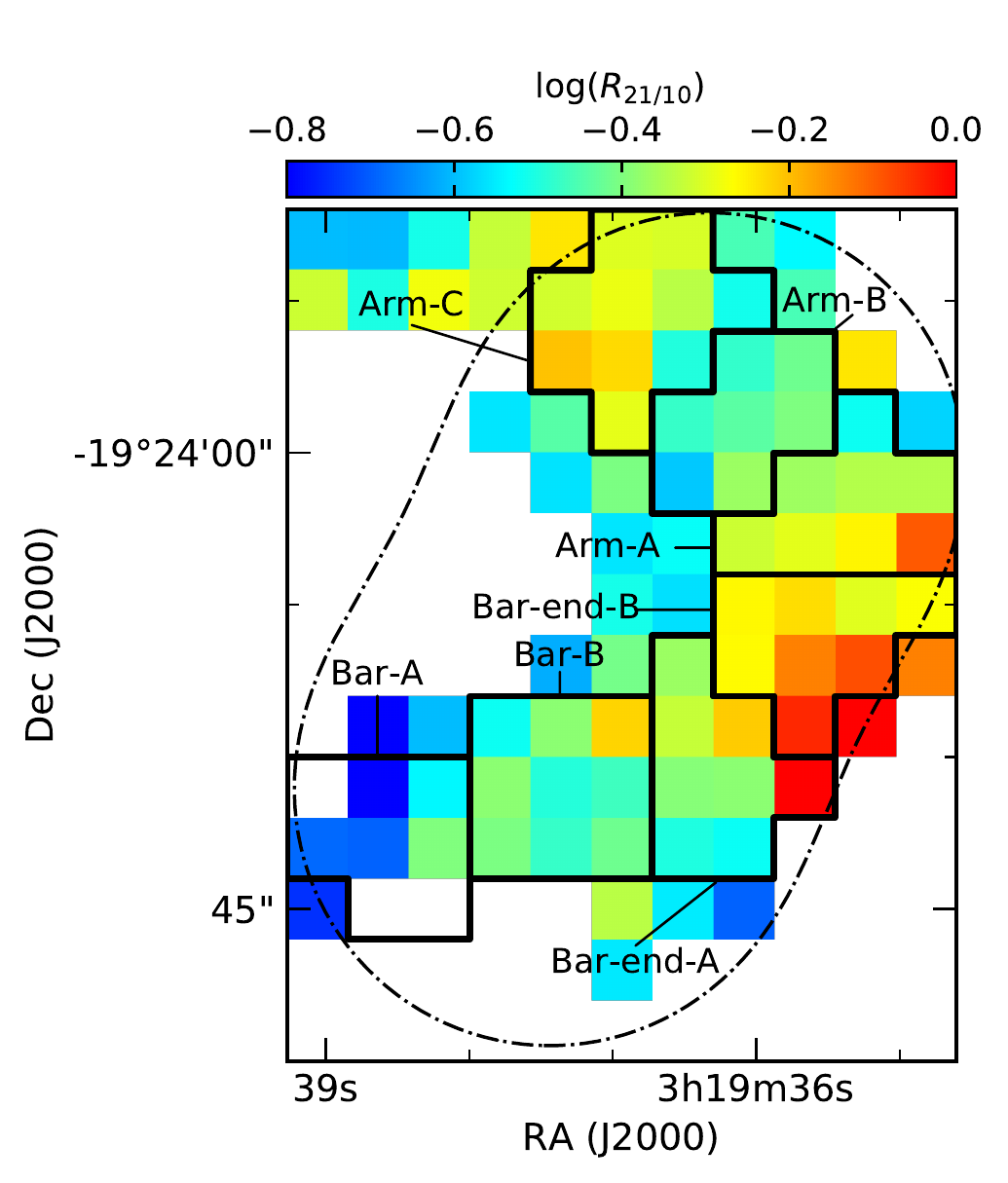}
    \caption{Spatial distribution of $R_{21/10}$.
    All lines are the same as Fig.~\ref{fig:Result_ICO10_ALMA}.  }
 \label{fig:Result R21}
\end{figure}

\begin{figure}
\begin{center}
	\includegraphics[width=75mm]{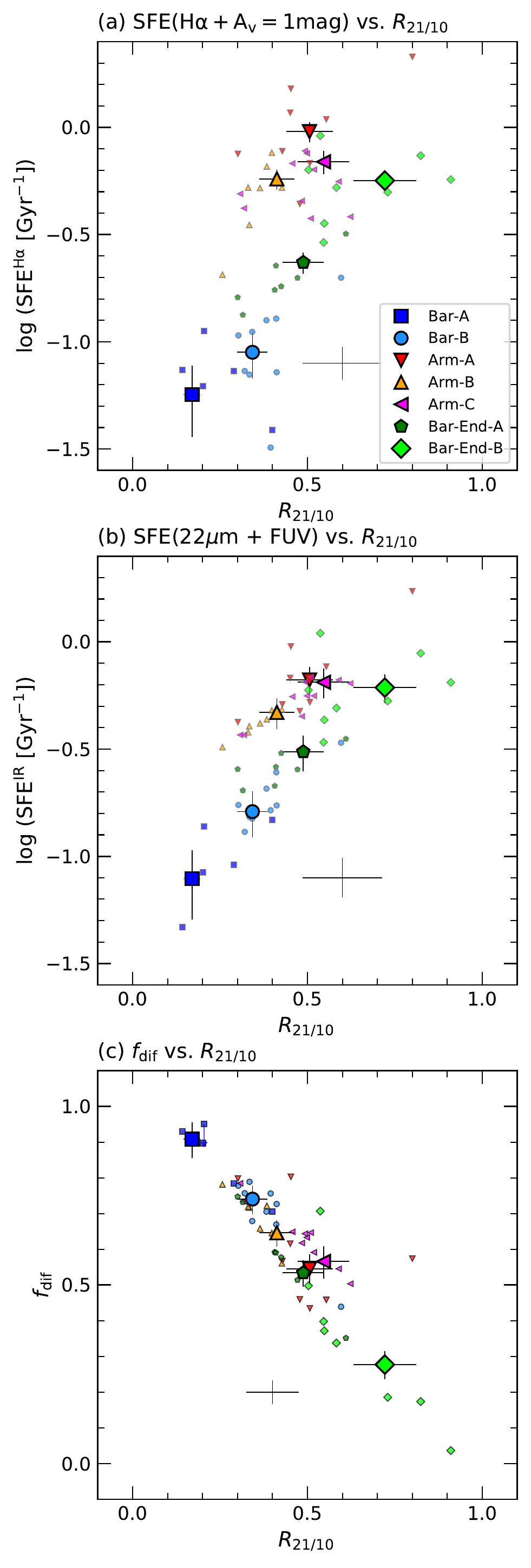}
    \caption{ (a) Correlation of $R_{21/10}$ with SFE. SFR is derived from the H$\alpha$ emission.  The small and large symbols represent the value in individual 6.0 arcsec pixels and the mean value in each environment, respectively.
    The typical uncertainty for each 6.0 arcsec pixel is shown as a black cross.
    (b) Same as panel (a), but SFR is derived from FUV and IR.
    (c) Correlation of $R_{21/10}$ with $f_{\rm dif}$.
     }
 \label{fig:Result_ALMAFoV_R21_SFE}
 \end{center}
\end{figure}

\section{Discussion}\label{sec:discussion}
\subsection{Trends with $R_{21/10}$} \label{sec: Result SFE vs. R21}

As described in Section~\ref{sec: Result SFE vs. Recovery fraction}, SFE decreases towards high $f_{\rm dif}$ in NGC~1300, which suggests that a large amount of diffuse molecular gases makes SFE low in the bar region.
In this section, to further investigate the physical condition of molecular gases, we measure the $^{12}$CO($2-1$)/$^{12}$CO($1-0$) line ratio, $R_{21/10}$, in NGC~1300 and examine the relation between $R_{21/10}$ and other physical parameters.
$R_{21/10}$ has been known for the measure to know the physical condition of molecular gases. Systematic variation of $R_{21/10}$ have reported in molecular clouds in the Milky Way \citep[e.g.][]{Sakamoto1994ApJ,Sakamoto1995ApJS}  and in galaxy of M51 \citep{Koda2012ApJ761} and M83 \citep{Koda2020arXiv200111043K}. 
The systematic variations of $R_{21/10}$ can be interpreted as changes of H$_2$ gas density and/or temperature \citep{Koda2012ApJ761}.

We derived $R_{21/10}$ from total $^{12}$CO($1-0$) flux obtained from the NRO 45-m (Section~\ref{sec: NRO CO(1-0)}) and total $^{12}$CO($2-1$) flux obtained from ALMA archival data (Section~\ref{sec: ALMA CO(2-1)}) as
\begin{equation}
    R_{21/10} = \frac{I_{\rm ^{12}CO(2-1)}^{\rm tot}}{I_{\rm ^{12}CO(1-0)}^{\rm tot}}.
\end{equation}
Fig.~\ref{fig:Result R21} shows the resultant spatial distribution of  $R_{21/10}$ in NGC~1300. We display the pixels where both the $I_{\rm ^{12} CO(1-0)}^{\rm tot}$ and $I_{\rm ^{12}CO(2-1)}^{\rm tot}$ were significantly detected. 
The uncertainties of  $R_{21/10}$ are mainly propagated from the $I_{\rm ^{12} CO(1-0)}^{\rm tot}$. 
We find that systematic variation of $R_{21/10}$:
In {\it bar-end-B}, the $R_{21/10}$ is the highest ($\sim 0.7$), followed by {\it bar-end-A}, {\it arm-A}, {\it arm-B} and {\it arm-C} ($\sim 0.4 -0.6$). In {\it bar-A} and {\it bar-B}, the $R_{21/10}$ is the lowest ($0.2 - 0.4$). This tendency is the opposite of $f_{\rm dif}$ (Section~\ref{sec: Diffuse gas fraction}).

Fig.~\ref{fig:Result_ALMAFoV_R21_SFE}(a) and (b) show the correlations between $R_{21/10}$ with SFE. Large symbols show the mean values in the environments, which are listed in Table \ref{tab:Stacking Analysis ALMAFoV}.
We find that the SFE decreases with decreasing $R_{21/10}$ for the large symbols.
This clear correlation have been reported in M51 \citep{Koda2012ApJ761}.
We find that a tight correlation of $R_{21/10}$ and $f_{\rm dif}$ as shown in Fig.~\ref{fig:Result_ALMAFoV_R21_SFE}(c).
The lower $R_{21/10}$ and higher $f_{\rm dif}$ are seen in  the bar regions where SFE is also low, while higher $R_{21/10}$ and lower $f_{\rm dif}$ are seen in active star forming regions, or arm and bar-end regions.

According to the non-LTE analysis by \citet{Koda2012ApJ761}, the higher $R_{21/10}$ in the star-forming regions implies that the gas may be warmer and/or denser. Thus, high $R_{21/10}$ and low $f_{\rm dif}$ in the star-forming regions (i.e. arm and bar-end) in NGC~1300 are interpreted as follows; most of the molecular gas is composed of molecular clouds there, resulting in a higher density of the molecular gas. In these regions, OB stars are associated with in the molecular clouds, which heat the molecular gas (i.e. stellar feedback).
On the other hand, in the bar regions, the gas density may be low because diffuse gas would dominate shown as high $f_{\rm dif}$. Because there is no massive SF, stellar feedback would not work there and gas temperature may be low. As a result of these, $R_{21/10}$ is considered to be low in the bar regions.

The $^{13}$CO($1-0$)/$^{12}$CO($1-0$) line ratio, $R_{13/12}$, is also a tool to investigate the physical condition of molecular gases.
It is reported that $R_{13/12}$ in the bar regions is lower than that in the arm regions \citep[e.g.][]{Watanabe2011,MorokumaMatusi2015PASJ,Yajima:2019do}. \citet{Watanabe2011} find the low $R_{13/12}$ in the bar regions, which indicates the presence of diffuse gas as a result of the strong streaming motion. Further, combination of $R_{21/10}$ and $R_{13/12}$ can be used to limit the gas density and temperature \citep[e.g.][]{Muraoka:2016ip}. Due to the non-detection of $^{13}$CO($1-0$) in NGC~1300 with the NRO 45-m, however, we cannot examine the variation of $R_{13/12}$ with environments, which remains for the next study.
In table \ref{tab:Stacking Analysis ALMAFoV}, we show the upper limit of $R_{13/12}$ using stacked spectra. Here, we estimate the 3$\sigma$ upper limit of $I_{\rm ^{13}CO(1-0)}$ as $3 \sigma_{\rm rms}^{13} \sqrt{{\rm FWHM}_{\rm ^{12}CO(1-0)} \Delta V}$, where $\sigma_{\rm rms}^{13}$ is the rms noise level of stacked $^{13}$CO($1-0$) spectrum, ${\rm FWHM}_{\rm ^{12}CO(1-0)}$ is the velocity width of the 
$^{12}$CO($1-0$) spectrum, and $\Delta V$ is the channel width of 10 $\rm km~s^{-1}$.

\begin{figure*}
\begin{center}
	\includegraphics[width=140mm]{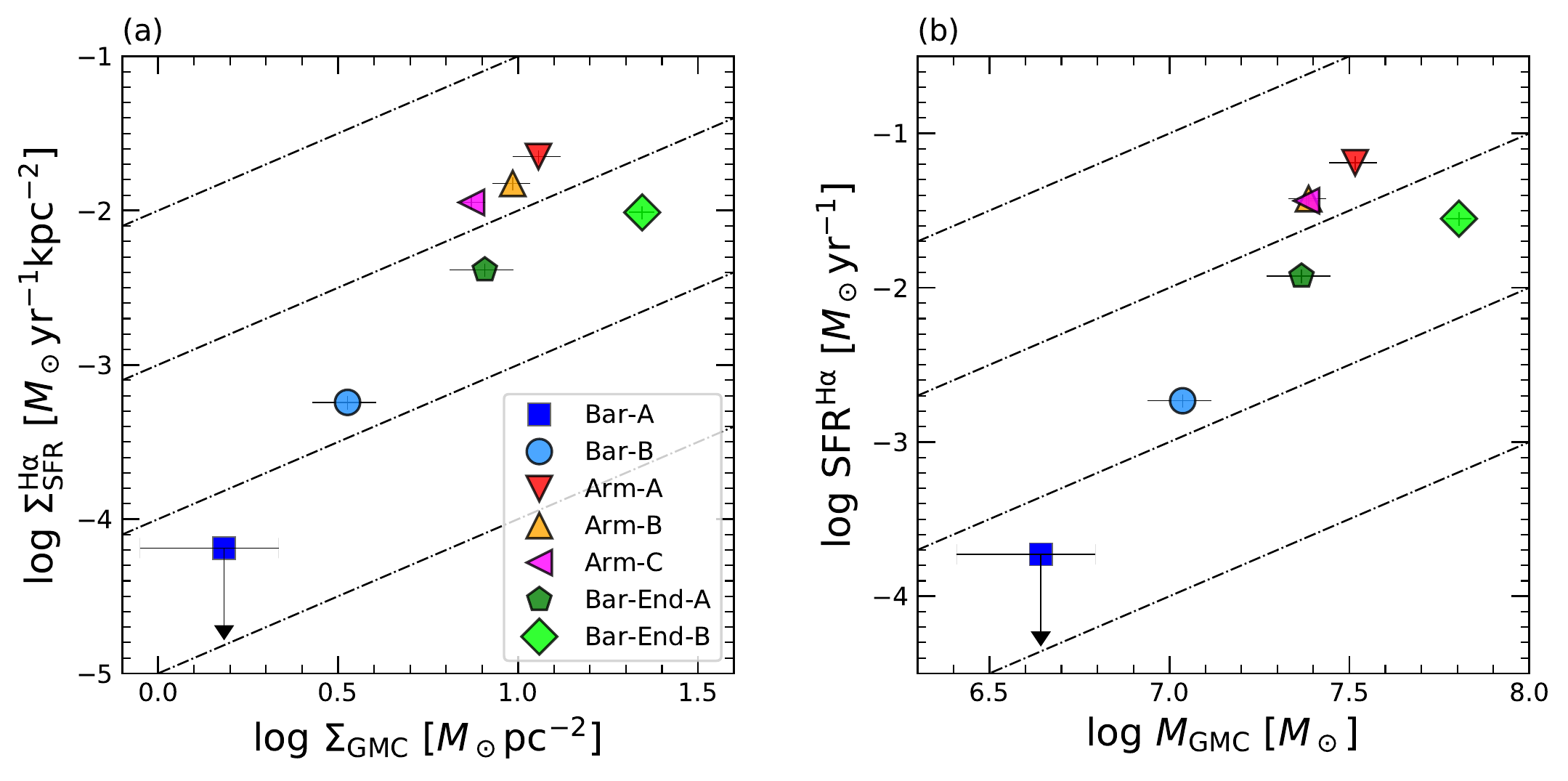}
    \caption{(a) The SFEs of molecular gas excluding diffuse components (i.e. molecular gas organized inside GMCs) in NGC~1300: SFR surface density derived from H$\alpha$ vs. molecular gas surface density derived from the GMC catalog \citep{Maeda2020MNRAS} in each environment (see text). 
    The black dash--dotted lines show the constant SFE of $10^1$, $10^0$, $10^{-1}$, and  $10^{-2}~\rm Gyr^{-1}$.
    (b) Same as panel (a), but used total SFR and total molecular gas mass.}
 \label{fig:SFR_vs_total_GMC}
 \end{center}
\end{figure*}

\begin{table*}
 \caption{SFE of the molecular gas excluding diffuse components.}
 \label{tab:GMC SFE}
 \begin{tabular}{llccccccc}
  \hline
& & Bar-A & Bar-B & Arm1-A & Arm1-B & Arm1-C & Bar-end-A & Bar-end-B\\
\hline
${\rm SFE}_{\rm GMC}$ &($\rm Gyr^{-1}$) &$<0.04$&$0.17 \pm 0.04$&$1.97\pm0.36$&$1.55\pm0.24$&$1.52\pm0.20$&$0.51\pm0.11$&$0.44\pm0.06$ \\
\hline
 \end{tabular}
\end{table*}

\subsection{The cause for the low SFE in the bar region}
\label{The cause for the low SFE in the bar region}

As described in Section~\ref{sec: Result SFE}, SFE in the bar regions is clearly suppressed in the strongly barred galaxy of NGC~1300. 
Our analysis shows the SFE decreases towards high $f_{\rm dif}$ and low $R_{21/10}$. 
These correlations support the idea that
the diffuse gas contributes significantly to the $\Sigma_{\rm H_2}^{\rm tot}$ used in the SFE calculation, which makes the SFE apparently low in the bar regions. 
If the environmental dependence of SFEs comes only from the difference of the fraction of diffuse gas, we expect that SFEs will be constant if we recalculate them excluding the diffuse molecular gas in the estimates of total gas mass.
To check this, we examine the following SFE defined as
\begin{equation}
    \left( \frac{\rm SFE_{\rm GMC}}{\rm Gyr^{-1}} \right) = 10^3
    \left( \frac{\Sigma_{\rm SFR}^{\rm H\alpha}}{M_\odot~\rm yr^{-1}~kpc^{-2}} \right)
    \left( \frac{\Sigma_{\rm GMC}}{M_\odot~\rm pc^{-2} } \right)^{-1},
\end{equation}
where $\Sigma_{\rm SFR}^{\rm H\alpha}$ is the total SFR surface density derived from H$\alpha$ emission by assuming $A_V = 1.0$~mag and $\Sigma_{\rm GMC}$ is the surface density of the molecular gas organized inside GMCs in each environment.
Here, the $\Sigma_{\rm SFR}^{\rm H\alpha}$ is derived from the original (not deconvolved and regrided) H$\alpha$ image. We first measured the total SFR in the environment, and then divided it by the area of the environment.
Because of the absence of H\textsc{ii} regions, the upper limit of the $\Sigma_{\rm SFR}^{\rm H\alpha}$ in {\it bar-A} is calculated using the SFR of an H\textsc{ii} region with a size of 50~pc whose H$\alpha$ flux in each pixel is 3 times of the noise level.
The $\Sigma_{\rm GMC}$ is derived using the GMC catalogue in NGC~1300 obtained by \citet{Maeda2020MNRAS}. Using 3D clumps finding algorithm CPROPS \citep{RosolowskyLeroy}, \citet{Maeda2020MNRAS} identified and characterized 233 GMCs with the mass completeness limit of $2.0 \times 10^5~M_\odot$. 
We first measured the total mass of the GMCs in the environments, and then divided it by the area of the environment. Note that $\Sigma_{\rm GMC}$ is not a gas surface density of  individual GMC.

Fig. \ref{fig:SFR_vs_total_GMC}(a) shows the Kennicutt-Schmidt diagram using $\Sigma_{\rm SFR}^{\rm H\alpha}$ and $\Sigma_{\rm GMC}$. 
We present the same diagram, but using the total SFR and total mass of the GMCs in Fig. \ref{fig:SFR_vs_total_GMC}(b) to avoid the uncertainty of the area. 
The $\rm SFE_{GMC}$ in each environment is listed in Table \ref{tab:GMC SFE}. 
The significant variations of $\rm SFE_{GMC}$ with environments are still seen: the $\rm SFE_{GMC}$ in the arm regions is the highest ($\sim 1.6~\rm Gyr^{-1}$), followed by the bar-end ($\sim 0.5~\rm Gyr^{-1}$) and the bar regions ($0.17~\rm Gyr^{-1}$ in {\it bar-B} and $< 0.04~\rm Gyr^{-1}$ in {\it bar-A}). 
These results can easily be expected from the fact that the GMCs exists in the bar regions where the prominent H\textsc{ii} region does not exist.
It is unveiled that the SFE excluding diffuse components is significantly suppressed in the bar regions,  which suggests the presence of other causes, in addition to a large amount of the diffuse molecular gases.
The fact that the $\rm SFE_{GMC}$  is different by a factor of 3 between the arm and bar-end regions also may reflect differences of the SF process of GMCs (see Section~\ref{sec: Result SFE vs. Recovery fraction}).

Why is the SFE excluding the diffuse molecular gas suppressed in the bar region of NGC~1300?
The cause is discussed in detail in \citet{Maeda2020MNRAS} and \citet{Fujimoto2020MNRAS}. Thus we briefly discuss here.
There are two possibilities for the cause.
One is that GMCs in the bar regions may be gravitatinally unbound \citep[e.g.][]{Sorai2012PASJ,Meidt2013ApJ,Nimori2013MNRAS}. However, this scenario is unlikely to be a main process of the SF suppression of the GMCs in NGC~1300;
\citet{Maeda2020MNRAS} found that there is no significant variations in  virial parameters of the GMCs among the bar, arm and bar-end regions using the GMCs catalogue in NGC~1300.
A hydrodynamical simulation modeled NGC~1300 by \citet{Fujimoto2020MNRAS} showed qualitatively similar result. These results suggest that the SF suppression in the bar can not be explained by a systematic difference of virial parameter.

Another possibility is that the fast CCCs may occur in the bar regions. It is thought that a fast CCC in the bars shortens a gas accretion phase of the cloud cores formed, leading to suppression of core growth and massive SF \citep[][]{Fujimoto:2014kh,Takahira:2014cq,Takahira:2018gx}. \citet{Fujimoto2020MNRAS} showed that the collision speed in the bar regions is significantly faster than the other regions. This result suggests the fast CCC is the physical mechanism which causes the lack of massive SF. Summarizing the discussion above, the low values of the SFE in the bar regions of NGC~1300 appear to be caused not only by the presence of a large amount of  diffuse molecular gases but also other mechanisms such as fast cloud--cloud  collisions among GMCs.

Why does a large amount of diffuse molecular gases exist in the bar regions? The answer is unclear. If
molecular clouds form from surrounding diffuse gases, one possibility is that efficiency of conversion from diffuse gases to molecular clouds is lower in the bar region. For example, a strong shock and/or shear along the bar may prevent forming GMCs from diffuse gases \citep[e.g.][]{Tubbs1982ApJ,Athanassoula1992MNRAS,ReynaudDownes1998A&A}.
Some observations reported the diffuse molecular gases constitute a thick disc
\citep[e.g.][]{Garcia-Burillo1992A&A,Sofue1993PASJ,Pety:2013fw}.
\citet{Pety:2013fw} estimated between 2 and 20 percent of the total molecular gas mass to be at galactic heights larger than 400 pc. Some simulations of the disc galaxies suggest that the cold molecular gas from the disc to the halo is caused by the feedback of massive SF \citep[e.g.][]{Wada2008ApJ.675,Dobbs2011MNRAS.417}.
However, it would be unlikely that the large amount of diffuse molecular gases forms the thick molecular gas disc in the bar regions of NGC~1300 because of the absence of the massive SF.

\section{Summary}\label{sec:summary}
In order to understand the cause for the low SFE in the bar regions, we test the scenario that the diffuse gas greatly contributes to the $\Sigma_{\rm H_2}$. We measured the diffuse molecular gas fraction ($f_{\rm dif}$) in the strongly barred galaxy NGC~1300, which is one of the suitable laboratories because the absence of the SF is clearly seen in the bar regions.
The $f_{\rm dif}$ is measured using $^{12}$CO($1-0$) observations data obtained with the single-dish NRO 45-m and with the ALMA 12-m array, which has no sensitivity on diffuse (extended) molecular gases due to the lack of ACA. Using SFR derived from archival H$\alpha$, FUV and IR data, we investigated the correlations $f_{\rm dif}$ with SFE in NGC~1300. The main results are as follows:

\begin{enumerate}
    \item The $\Sigma_{\rm mol}^{\rm tot}$ in bar, arm, and bar-end regions of NGC~1300 ranges from 10 to 15 $\rm K~km~s^{-1}$, which indicates that the molecular gases do exist in the strong bars with no clear H\textsc{ii} regions (Fig.~\ref{fig:Result_ICO10_NRO}). The $\Sigma_{\rm mol}^{\rm tot}$ in the bar regions is lower than those in other nearby barred galaxies (Section~\ref{sec: conversion to Smol}).
    
    \item We derived SFR from H$\alpha$ emission assuming $A_V = 1.0~\rm mag$ and hybrid tracers (FUV + IR). The magnitude of systematic differences in both SFRs rarely exceeds a factor of two. While the $\Sigma_{\rm SFR}$ is $10^{-2.7}-10^{-2.1}~M_\odot~\rm yr^{-1}~kpc^{-2}$ in arm and bar-end regions, $\Sigma_{\rm SFR}$ is mostly lower than $10^{-3}~M_\odot~\rm yr^{-1}~kpc^{-2}$ in bar regions (Section~\ref{sec: conversion to Smol} and Fig.~\ref{fig:Result_SFR_map_and_SFR_comparison}).
    
    \item SFEs in the bar regions are  $5 - 10$ times lower than those in the arm and bar-end regions despite the  $\Sigma_{\rm H_2}$ is comparable between the both regions in NGC~1300: $(0.06-0.16)~\rm Gyr^{-1}$ in the bar regions and $(0.47-0.96)~\rm Gyr^{-1}$ in the arm and bar-end regions. Our result indicates that the SF in the bar regions of the strongly barred galaxies is further suppressed than in the intermediate-type barred galaxies (Section~\ref{sec: Result SFE}, Fig.~\ref{fig:Result_fre_SFE_map_in_ALMAFoV}(a) and (b), and Fig.~\ref{fig:Result_ALMAFoV_KS_fre_SFE}(a) and (b)).
    
    \item We found difference of the $f_{\rm dif}$ among the environments: the $f_{\rm dif}$ is  $0.74 - 0.91$ in the bar regions and $0.28 - 0.65$ in the arm and bar-end regions. This indicates most of the molecular gas in the bar region exists as the diffuse gas with FWHM $\gtrapprox 700~\rm pc$
    (Section~\ref{sec: Diffuse gas fraction} and Fig.~\ref{fig:Result_fre_SFE_map_in_ALMAFoV}(c)).
    
    \item We found the SFE decreases towards high  $f_{\rm dif}$, which suggests that the presence of a large amount of the diffuse molecular gases makes the SFE low in appearance (Section~\ref{sec: Result SFE vs. Recovery fraction} and  Fig.~\ref{fig:Result_ALMAFoV_KS_fre_SFE}(c) and (d)).
    
    \item We found a tight negative correlation between  the line ratio of $R_{21/10}$ and the $f_{\rm dif}$. This result supports the idea that  a large amount of diffuse molecular gases makes the SF low and suggests low gas density and/or temperature in the bar regions.
    (Section~\ref{sec: Result SFE vs. R21} and Fig.~\ref{fig:Result_ALMAFoV_R21_SFE}).
    
    \item The SFEs of molecular gas excluding diffuse components (i.e. molecular gas organized inside GMCs) in the bar regions are significantly lower than those in the arm and bar-end regions, which suggests the presence of  other causes for the suppression besides the large amount of the diffuse molecular gases. Based on the studies on GMC properties in NGC~1300 (\citealt{Maeda2020MNRAS} and \citealt{Fujimoto2020MNRAS}), the low SFE may be caused not only by a large amount of  diffuse molecular gases but  also other mechanisms such as fast cloud--cloud  collisions among GMCs  (Section~\ref{The cause for the low SFE in the bar region}).
\end{enumerate}

\section*{Acknowledgements}
We would like to thank the referee for useful comments.
We are grateful to K. Nakanishi, H. Kaneko, G. Kim, and
the staff at the ALMA Regional Center and the staff at the Nobeyama Radio Observatory (NRO) for their help in observations and data reduction. FM is supported by Research Fellowship for Young Scientists from the Japan Society of the Promotion of Science (JSPS). KO is supported by JSPS KAKENHI Grant Numbers JP16K05294 and JP19K03928.
AH is funded by the JSPS KAKENHI Grant Number JP19K03923.
The Nobeyama 45-m radio telescope is operated by NRO, a branch of National Astronomical Observatory of Japan (NAOJ).
This paper makes use of the following ALMA data: ADS/JAO.ALMA\#2017.1.00248.S.
and \#2015.1.00925.S. ALMA is a partnership of ESO (representing its member states), NSF (USA) and NINS (Japan), together with NRC (Canada), MOST and ASIAA (Taiwan), and KASI (Republic of Korea), in cooperation with the Republic of Chile. The Joint ALMA Observatory is operated by ESO, AUI/NRAO and NAOJ.
Data analysis was in part carried out on the Multi-wavelength Data Analysis System operated by the Astronomy Data Center (ADC), NAOJ.

\bibliographystyle{mnras}
\bibliography{Reference} 

\appendix

\section{Supplementary results} \label{apx}
\begin{figure*}
	\includegraphics[width=\hsize]{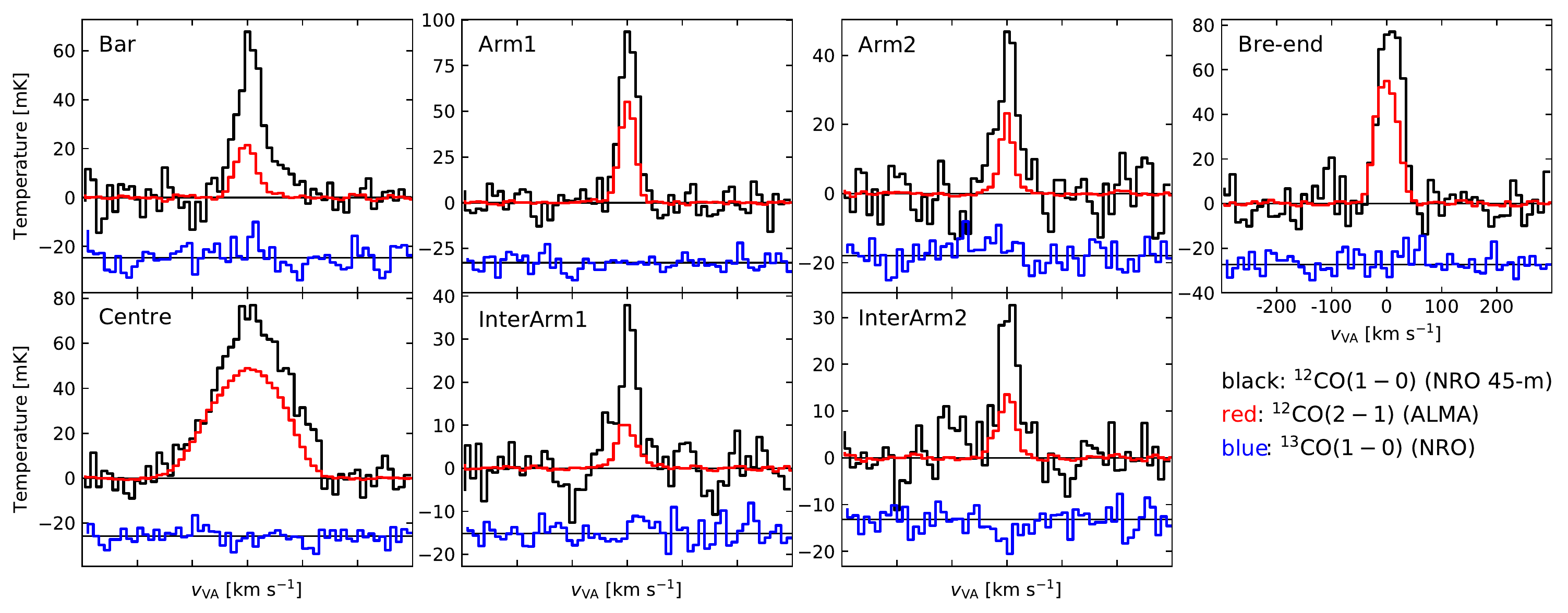}
    \caption{Stacked CO spectra for each environment in NGC~1300. The black, red, and blue lines indicate $^{12}$CO($1-0$) emission obtained from the NRO 45-m, $^{12}$CO($2-1$) emission obtained from ALMA, and  $^{13}$CO($1-0$) emission obtained from the NRO 45-m, respectively.
    For visualization purpose, base line of $^{13}$CO($1-0$) spectra is shifted.}
 \label{fig_apx:Result_stacking_analysis_NROFoV}
\end{figure*}

\begin{figure*}
	\includegraphics[width=160mm]{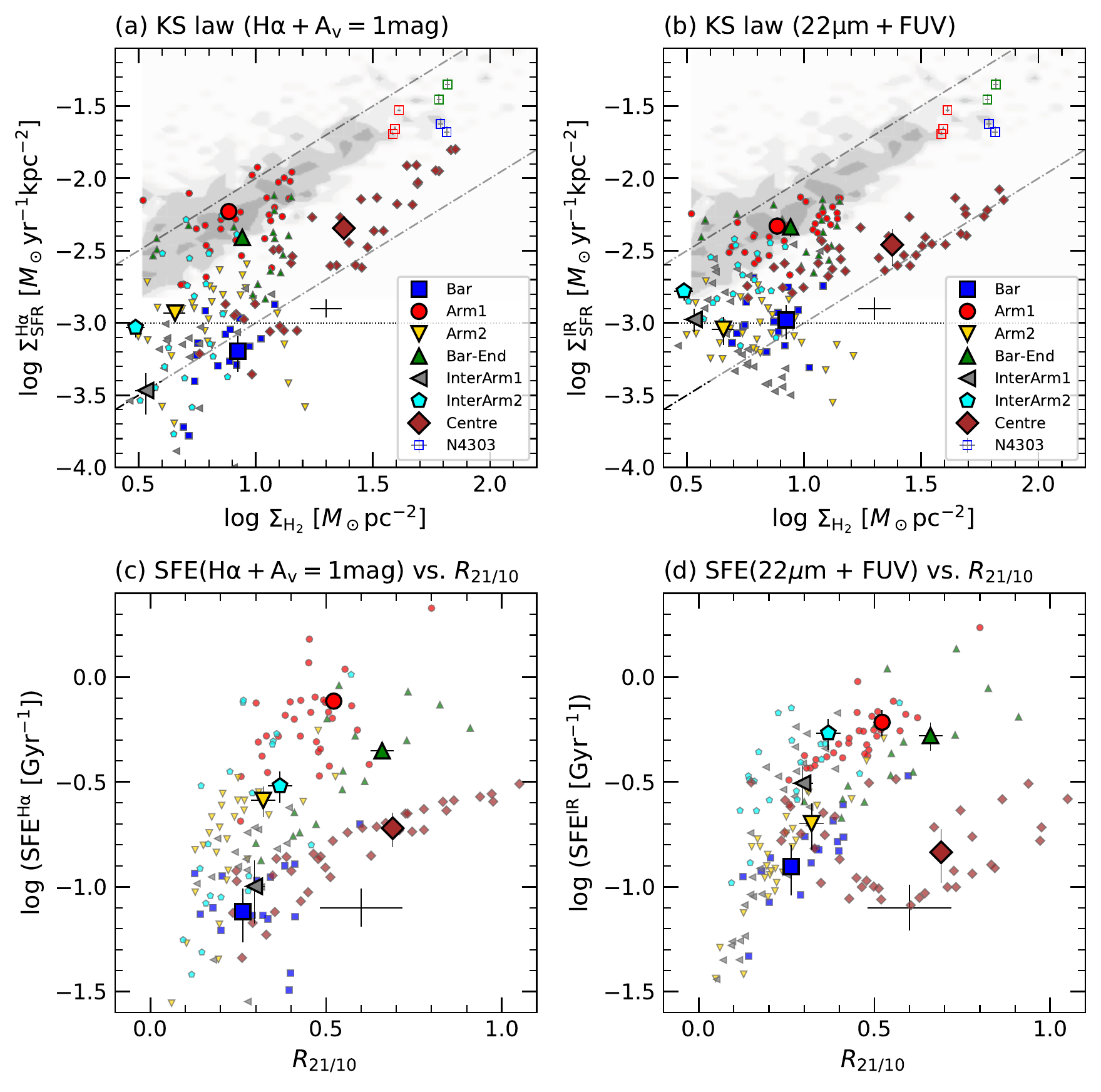}
    \caption{(a) Molecular Kennicutt-Schmidt relation of NGC~1300 for NRO FoV. SFR is derived from the H$\alpha$ emission. 
    The small and large symbols represent the value in individual 6.0 arcsec pixels and the mean value in each environment, respectively.
    We display the pixels where $^{12}$CO($1-0$) emission observed with the NRO 45-m was significantly detected.
    Typical uncertainty for each 6.0 arcsec pixel is shown as a black cross.
    The grey scale represents the result by \citet{Bigiel2008AJ}. 
    The open squares represents the result in NGC~4303 by \citet{Yajima:2019do} (blue: bar, red: arm, and green: bar-end). 
    The grey dash--dotted lines show the constant SFE of $1.0$ and $0.1~\rm Gyr^{-1}$.
    The horizontal dotted line represents  $\Sigma_{\rm SFR} = 10^{-3}~M_\odot~\rm yr^{-1}~kpc^{-2}$.
    (b) Same as panel (a), but SFR is derived from FUV and IR.
    (c) Correlation of $R_{21/10}$ with SFE for NRO FoV. SFR is derived from the H$\alpha$ emission.  The small and large symbols represent the value in individual 6.0 arcsec pixels and the mean value in each environment, respectively.
    The typical uncertainty for each 6.0 arcsec pixel is shown as a black cross.
    (d) Same as panel (c), but SFR is derived from FUV and IR. }
 \label{fig_apx:Result_NROFoV_KS_R21}
\end{figure*}

Here we present the results in NRO FoV as  supplementary results, because the $f_{\rm dif}$ and the relations between $f_{\rm dif}$ and other parameters cannot be measured outside ALMA FoV.
Fig.~\ref{fig_apx:Result_stacking_analysis_NROFoV} shows stacked CO spectra in each environment defined as Fig.~\ref{fig:definition_of_environments}(a). The mean values of physical parameters in each environment are listed in Table \ref{tab_apx:Stacking Analysis NROFoV}.
In {\it inter-arm1} and {\it inter-arm2}, the $\Sigma_{\rm ^{12}CO(1-0)}^{\rm tot}$ are the lowest ($\sim 4 -5~M_\odot~\rm pc^{-2}$).
In {\it centre}, the $\Sigma_{\rm ^{12}CO(1-0)}^{\rm tot}$ and $R_{21/10}$ are the highest ($32.2~M_\odot~\rm pc^{-2}$ and $0.69$). However, SFE in {\it centre} is about $3-4$ times lower than those in {\it arm} and {\it bar-end}. This result is clearly seen in K-S law in Fig.~\ref{fig_apx:Result_NROFoV_KS_R21}(a) and (b), which indicates that SF in centre of NGC~1300 is suppressed.
However, the adopted $\alpha_{\rm CO}=  4.4~M_\odot (\rm K~km~s^{-1}~pc^2)^{-1}$ may not be appropriate in the centre region. In central regions of disc galaxies $\alpha_{\rm CO}$ is lower than by a factor of $2-3$ (some nearby galaxies; \citealt{Nakai_Kuno1995PASJ}; \citealt{Regan2000ApJ}, the Milky Way; \citealt{Dahmen1998A&A}). 
Fig.~\ref{fig_apx:Result_NROFoV_KS_R21}(c) and (d) show the correlation of $R_{21/10}$ with SFE for NRO FoV.
The 6.0 arcsec pixels in {\it arm2}, {\it inter-arm1}, and {\it inter-arm2} lie on the tendency we find in bar, arm, and bar-end regions (Fig.~\ref{fig:Result_ALMAFoV_R21_SFE}(a) and (b)).
Pixels in {\it centre} deviated from the tendency; but this may be due to the systematic variation of $\alpha_{\rm CO}$.

\begin{table*}
 \caption{Mean values of physical parameters for NRO FoV.}
 \label{tab_apx:Stacking Analysis NROFoV}
 \begin{tabular}{llccccccc}
  \hline
& & Bar & Arm1 & Arm2 & Bar-end & Centre & Inter-arm1 & Inter-arm2\\
\hline
$\Sigma_{\rm SFR}^{\rm H\alpha}$ &($10^{-3}~M_\odot \rm~yr^{-1}~kpc^{-2}$)   &$0.6\pm0.2$&$5.9\pm0.4$&$1.2\pm0.2$&$3.9\pm0.3$&$4.5\pm0.8$&$0.34\pm0.1$&$0.9\pm0.1$\\
$\Sigma_{\rm SFR}^{\rm IR}$ &($10^{-3}~M_\odot \rm~yr^{-1}~kpc^{-2}$)  &$1.0\pm0.3$&$4.7\pm0.6$&$0.9\pm0.2$&$4.6\pm0.7$&$3.5\pm1.0$&$1.0\pm0.2$&$1.7\pm0.2$\\
$I_{\rm ^{12}CO(1-0)}^{\rm tot}$ &  ($\rm K~km~s^{-1}$)&$4.1\pm0.2$&$3.7\pm0.2$&$2.2\pm0.2$&$4.2\pm0.2$&$11.5\pm0.3$&$1.6\pm0.2$&$1.5\pm0.1$ \\
$I_{\rm ^{12}CO(2-1)}^{\rm tot}$ & ($\rm K~km~s^{-1}$)&$1.07\pm0.02$&$1.93\pm0.02$&$0.70\pm0.01$&$2.79\pm0.02$&$7.92\pm0.02$&$0.49\pm0.01$&$0.55\pm0.01$ \\
$\Sigma_{\rm ^{12}CO(1-0)}^{\rm tot}$ & ($M_\odot \rm~pc^{-2}$)  &$11.4\pm0.6$&$10.4\pm0.5$&$6.2\pm0.7$&$11.9\pm0.6$&$32.3\pm0.8$&$4.6\pm0.4$&$4.2\pm0.4$\\
${\rm SFE}^{\rm H\alpha}$ &($\rm Gyr^{-1}$)  &$0.08\pm0.02$&$0.77\pm0.07$&$0.26\pm0.04$&$0.45\pm0.04$&$0.19\pm0.04$&$0.10\pm0.03$&$0.30\pm0.05$\\
${\rm SFE}^{\rm IR}$ &($\rm Gyr^{-1}$) &$0.12\pm0.03$&$0.61\pm0.09$&$0.20\pm0.05$&$0.53\pm0.08$&$0.15\pm0.04$&$0.31\pm0.06$&$0.54\pm0.09$\\
$R_{21/10}$ &  &$0.26\pm0.02$&$0.52\pm0.03$&$0.32\pm0.03$&$0.66\pm0.03$&$0.69\pm0.02$&$0.30\pm0.03$&$0.37\pm0.03$\\
$R_{\rm 13/12}$ & &$<0.09$&$<0.07$&$<0.12$&$<0.06$&$<0.03$&$<0.11$&$<0.10$\\
\hline
 \end{tabular}
\end{table*}

\bsp	
\label{lastpage}
\end{document}